%% file: final_manuscript_AA_2018_33767.tex
\documentclass{aa}
\usepackage{graphicx}
\usepackage[varg]{txfonts}
\usepackage{natbib}
\bibpunct{(}{)}{;}{a}{}{,} 
\usepackage{enumitem}
\usepackage{amsmath}	
\usepackage{amssymb}	
\usepackage{booktabs}
\usepackage{longtable}
\newcommand{\kms}{\mbox{km\ s}$^{-1}$}

\newcommand{\msun}{M_\odot}

\newcommand{\Omo}{\Omega_{{\rm m},0}}

\newcommand{\Olo}{\Omega_{\Lambda,0}}

\newcommand{\Pa}{P_{\rm dagn}}
\newcommand{\Pao}{P_{\rm dagn,spec}}
\newcommand{\Pac}{P_{\rm dagn}^{\rm pair}}
\newcommand{\Paoc}{P_{\rm dagn,spec}^{\rm pair}}
\newcommand\T{\rule{0pt}{2.6ex}}       
\newcommand\B{\rule[-1ex]{0pt}{0pt}} 

\begin{document}

\title{Intrinsic and observed dual AGN fractions from major mergers}


\titlerunning{Dual AGN fractions from major mergers}
\authorrunning{J.~M.\ Solanes et al.}
\author{J.~M.\ Solanes
  \inst{1,2}
  \and
  J.~D.\ Perea
  \inst{3}
  \and
  G.\ Valent\'{\i}--Rojas
  \inst{1}
  \and
  A. del Olmo
  \inst{3}
  \and
  I. M\'arquez
  \inst{3}
  \and
  C. Ramos Almeida
  \inst{4,5}
  \and
  J.~L.\ Tous
  \inst{1}}

\institute{Departament de F\'{\i}sica Qu\`antica i Astrof\'{\i}sica, Universitat de
  Barcelona. C.\ Mart\'{\i} i Franqu\`es, 1; E--08028~Barcelona, Spain
  \and
  Institut de Ci\`encies del Cosmos (ICCUB), Universitat de
  Barcelona. C.\ Mart\'{\i} i Franqu\`es, 1; E--08028~Barcelona, Spain
  \and
  Departamento de Astronom\'{\i}a Extragal\'actica, Instituto de
  Astrof\'{\i}sica de Andaluc\'{\i}a, IAA--CSIC. Glorieta de la
  Astronom\'\i a, s/n; E--18008~Granada, Spain
  \and
  Instituto de Astrof\'{\i}sica de Canarias, C.\ V\'{\i}a L\'actea, s/n; E--38205~La Laguna,
  Tenerife, Spain
  \and
  Departamento de Astrof\'{\i}sica, Universidad de La Laguna; E--38206~La Laguna, Tenerife, Spain
  }

\offprints{J.~M.\ Solanes, \email{jm.solanes@ub.edu}}

\date{Accepted by \aap.}


\abstract{A suite of 432 collisionless simulations of bound pairs of
  spiral galaxies with mass ratios 1:1 and 3:1, and global properties
  consistent with the $\Lambda$CDM paradigm, is used to test the
  conjecture that major mergers fuel the dual AGN (DAGN) of the local
  volume. Our analysis is based on the premise that the essential aspects of this  
  scenario can be captured by replacing the physics of
  the central black holes with restrictions on their
  relative separation in phase space. We introduce several estimates of the DAGN fraction 
  and infer predictions for the activity levels and resolution limits
  usually involved in surveys of these systems, assessing their 
  dependence on the parameters controlling the length of both 
  mergers and nuclear activity. Given a set of constraints, we find that the values adopted for some of the latter factors often condition the outcomes from individual experiments. 
  Still, the results do not reveal, in general, very tight correlations, 
  being the tendency of the frequencies normalized to the merger time 
  to anticorrelate with the orbital circularity the clearest effect. In agreement with other theoretical studies of DAGN in galaxy mergers, our simulations predict intrinsic abundances of these systems that range from $\sim$ a few to $15\%$ depending on the maximum level of nuclear activity achieved, the higher the bolometric luminosity, the lower the fraction. At the same time, we show that these probabilities are reduced by about an order of magnitude when they are filtered with the typical constraints applied by observational studies of the DAGN fraction at low redshift.  Seen as a whole, the results of the present work prove that the 
  consideration of the most common limitations involved in the detection 
  of close active pairs at optical wavelengths is sufficient by itself to reconcile the 
  intrinsic frequencies envisaged in a hierarchical universe with 
  the small fractions of double-peaked 
  narrow-line systems which are often reported at kpc-scales.}
  
\keywords{galaxies: active -- galaxies: interactions -- galaxies:
  nuclei -- methods: numerical}

\maketitle

\section{Introduction}\label{Introduction}

Within the current paradigm of galaxy evolution through hierarchical
structure formation the close pairs of active galactic nuclei, which at
kpc-scale separations are usually referred to as dual AGN (DAGN), are
widely believed to be the later stages of major mergers of galaxies
\citep[e.g.][]{DiM05,Hop06,CD11,Fu11,Kos12,BLN13,Ell13,Fan16,Sha16,Wei18},
although some works seem to suggest otherwise
\citep[e.g.][]{Cis11,Sch12,Her16,Vil17}. In binary mergers, as the
progenitor galaxies orbit around each other, they transfer angular
momentum to the dark matter (DM) and to each other through dynamical
friction and start to sink towards the center of the system. If both
members of the pair are gas-rich and massive enough to contain a
central supermassive black hole \citetext{SMBH; \citealt{KH13}}, the
gravitational torques generated on every close passage are expected to
drive substantial inflows of gas to the inner regions of the merging
objects, where it can be accreted into the central BH and ignite 
nuclear activity. As the merger develops, the level of activity will
tend to progressively achieve higher values, with its peak likely at
the galaxies' final approach, when the two SMBH rapidly in-spiral to
the center of the remnant and coalesce emitting the most powerful pulses 
of gravitational waves in the universe \citep{Min17}. Of course, this
idealized view does not prevent the nuclear activity from
being (alternatively) restricted to one of the members of the pair, or
triggered occasionally by interactions not necessarily leading to a
merger, as observed in luminous quasars and radio sources
\citep{RamA11,Bes12}, or by secular (internal) processes \citep{MMP95,MM08}. In any
event, the detection and abundance of DAGN over galaxy-wide scales
not only provides an essential test for the merger-driven scenario of these
systems, but has also important implications for hierarchical
structure formation theories, the growth and demographics of SMBH, the
accretion and feedback physics, and even for our understanding of
gravity.

Over the last decade a number of systematic studies of AGN pairs have
been conducted at different wavelengths. Many have relied upon optical
data, such as the searches for double-peaked narrow emission
lines in the spectroscopic galaxy sample of the Sloan Digital Sky
Survey \citetext{SDSS; e.g. \citealt{Wan09,Liu10,Com11,Ell11,Ros11,Ge12}}.
Yet there is no dearth of examples of studies based on observations at
other wavelengths, such as those carried out in the X-ray
\citep[][]{Kos12,Ten12}, mid-IR \citep[][]{Sat17}, or radio windows
\citetext{\citealt{Fu15}; additional references can be found in
  \citealt{RDK18}}. In not a few cases these investigations have
reported a frequency of DAGN at kpc scales surprisingly low according
to current observational constraints on the merger rate of galaxies,
even after taking into account selection effects
\citep[e.g.][]{Liu11}. Naive calculations that compare typical merging
timescales to expected AGN lifetimes suggest that, in accordance with
the plausible scenario described above, the fraction of DAGN in the
local universe should be roughly one order of magnitude larger than
the values $\lesssim 1\%$ typically observed
\citep[e.g.][]{FVD09}. Nevertheless, it must not be forgotten either that 
observational DAGN studies suffer from biases related to the
wavelength and methodology used to diagnose the dual BH activity, as well
as of, in many cases severe, incompleteness related to the design of
the surveys, the identification of mergers, and the difficulty in
resolving the pairs at small projected separations.

This work focuses precisely on the triggering and detectability of DAGN at kpc-scales 
in the nearby universe under the assumption that the dual nuclear activity 
takes place in bound pairs of similarly massive spiral galaxies 
\citep[e.g.][]{Ste16}. Our aim is to investigate the
contribution of the limitations associated with the standard
methodologies used for the detection of active SMBH pairs 
to the aforementioned conflict between theory and
observations. We are particularly interested in assessing the importance 
in the selection
of DAGN candidates of the constraints stemming from the observation of
double-peaked narrow emission lines in optical spectra, as well as
of the most common limitations related to the selection of close active companions around
previously detected single AGN. In order to do so, a large
subset of simulated collisions between MW-like galaxies included in
the massive suite of high-resolution isolated binary mergers recently
analyzed by \citet{SPV18} has been used. While these merger experiments represent
neatly and extensively the sort of gravitational encounters most
likely involved in DAGN activation, they lack a self-consistent
treatment of the complex physical processes that deliver the gas to
the nuclear BH of the interacting galaxies. To solve this deficiency 
we have adopted the strategy of encoding the physics of the 
gaseous component by means of the kinematics of the collisions, 
representing it through a series of thresholds defined in the 
six-dimensional phase space of the relative separation of the two 
progenitor galaxies\footnote{In line with this approximation, our experiments neglect the slight
reduction on the dynamical friction timescale that may result in
moderately wet mergers (i.e. with gas fractions on the order of
$10\%$) from the cooling of the gas, which acts to enhance the
central mass concentration of the galaxies and favors the sinking
of the secondary object onto the primary one \citetext{see
\citealt{Col14} and references therein}.}. In addition, the intergalactic 
separations adopted as a proxy for the different levels of BH activity 
have been complemented by several sets of constraints, 
both along the line of sight (LOS) and in the plane of the sky, intended to represent 
the limitations habitually present in the, photometric and spectroscopic, detection 
of AGN pairs.

The paper is laid out as follows. We first outline in
Section~\ref{experiments} the main characteristics and initial conditions 
of the major
merger simulations used for the present study. Section~\ref{DAGNmodel}
introduces the strategy devised to estimate from our experiments the
incidence of DAGN in the local universe. We then discuss in
Section~\ref{probabilities} a number of possible alternatives for the
calculation of this fraction and the different outcomes we obtain,
paying special attention to the assessment of the importance of
merger parameters. Finally, in Section~\ref{comparison}, we analyze
the validity of our estimates of the DAGN fraction from the
standpoint of both observational studies and theoretical data provided
by the latest state-of-the-art numerical simulations, while
Section~\ref{conclusions} summarizes our work and discusses the main
insights that emerge from it. Figures showing our results for DAGN
fractions in close pairs are described in the Appendix and included as
online only material. All probabilities and magnitudes inferred in the
present investigation have been calculated assuming a standard flat
$\Lambda$ Cold Dark Matter ($\Lambda$CDM) cosmology with
$H_0=70$~km~s$^{-1}$~Mpc$^{-1}$, $\Omo=0.3$ and $\Olo=0.7$.

\section{Numerical models of binary mergers}\label{experiments}

The runs used for the present investigation constitute the S+S subset 
of the suite of simulations of isolated binary galaxy mergers in bound 
orbits by \citeauthor{SPV18} \citetext{\citeyear{SPV18}; see this work for full details}. 
The orbital configuration of the mergers is defined 
from the initial orbital energy in dimensionless form, represented as usual
by the ratio 
\begin{equation}
r_{\rm circ,p}\equiv\frac{r_{\rm circ}({\mathcal E})}{R_{\rm p}}=-\frac{GM_{\rm p}M_{\rm s}}{2{\mathcal E}R_{\rm p}}\;,
\end{equation}
with $M_{\rm p}$ and $M_{\rm s}$, respectively, the virial masses of the primary and secondary progenitor galaxies, $R_{\rm p}$ the virial radius of the primary's halo, and
$r_{\rm circ}({\mathcal E})$ the radius of a circular orbit with the
same orbital energy ${\mathcal E}$, as well as from the initial orbital
circularity
\begin{equation}
\epsilon\equiv\frac{{\mathcal L}}{{\mathcal L}_{\rm circ}({\mathcal E})}=\sqrt{\frac{-2{\mathcal E}}{\mu}}\frac{\mathcal L}{GM_{\rm p}M_{\rm s}}\;,
\end{equation}
which acts as a dimensionless proxy of the orbital spin, 
${\mathcal L}=\mu rv_{\rm tan}$, with $v_{\rm tan}$ the tangential 
component of the time derivative of the intercentric separation $\vec{r}=\vec{r}_{\rm p}-\vec{r}_{\rm s}$ and $\mu=M_{\rm p}M_{\rm s}/(M_{\rm p}+M_{\rm s})$ the reduced mass of the system. The values 
chosen for these two quantities are representative of their
probability density functions (PDF) predicted by the currently favored
$\Lambda$CDM cosmological model. For $r_{\rm circ,p}$ two values 
are considered: $4/3$, which approximates the peak of the
heavily right-skewed orbital energy distribution found in the
cosmological simulations by \citet{McCa12}, and $2.0$, intended to
account for the relatively abundant more energetic orbits 
(we note that this second value of the orbital energy is applied 
only to equal-mass mergers without affecting the conclusions of this work). We also include collisions along three different orbital
trajectories defined by $\epsilon=0.20$,~$0.45$, and $0.70$.  These values are arranged
more or less equidistantly across the universal and heavily platykurtic PDF of the orbital circularity shown by galaxies in bound orbits
\citep[e.g.][]{Ben05,KB06,Jia08}, whose peak is at $\epsilon\sim 0.5$ (corresponding to an ellipticity $e\simeq 0.9$). Each initial orbital setup is combined with 
four different values of the dimensionless internal spin \citep{Pee69} of the progenitor galaxies that sample the main part of its similarly universal $P(\lambda)$ where the probability is highest \citep[e.g.][]{Sha06,Her07,Bry13}: $\lambda=0.00$, $0.02$, $0.04$ ($\sim$ median), and $0.06$, and which are assumed identical for both members of the pair. Finally, when defining the geometry of the encounters, we have also considered galaxies with different initial relative orientations. Since cosmological simulations suggest that the orientations of the spins of merging halos and of the orbital angular momentum are basically uncorrelated \citep{KB06}, we have put the focus on all those extreme configurations (twelve) that maximize/minimize the coupling between the internal spin vectors of galaxies, or rotation axes, and the orbital spin, and hence that maximize/minimize the duration of the mergers (see Table~\ref{halo_spins}).

Our model galaxies are made up of an extended spherical \citet*{NFW97} 
DM halo whose global properties (mass, spin and concentration) are used to set the
scalings of its central baryonic (stellar) core. 
The mass of the central luminous component of the
galaxies is taken equal to $5\%$ of their total mass and distributed
in the form of an exponential disc of stars surrounding a non-rotating
spherical \citet{Her90} stellar bulge. Two values are adopted for the
total mass ratio of the primary and secondary
progenitors, $\eta\equiv M_{\rm p}/M_{\rm s} =1$ and $3$, which correspond to the boundaries of the major merger range. For the largest
progenitors, intended to represent a $\sim 10^{12}\msun$ galaxy of
Hubble's Sb class, the bulge mass, $M_{\rm b}$, is taken equal to the
$25\%$ of the disc mass, $M_{\rm d}$, while the smallest
progenitors, which picture local Sc galaxies, have
$M_{\rm b}=0.1M_{\rm d}$ \citep{Gra01}. 

The largest galaxies are modeled using a total of $210,000$ particles, while
for the smaller objects this number is scaled by the
factor $1/\eta$. All experiments adopt a fifty-fifty split
in number between luminous and dark bodies. The Plummer equivalent
softening length for the luminous particles is set to 30 pc, while for
the more massive bodies (DM), the softening length is taken proportional
to the square root of their body mass, thereby ensuring the same
maximum interparticle gravitational force. Although the galaxy models
allow a single extra particle representing a SMBH to be placed right
at the center, the extent of the DAGN phenomenon has been simulated in
practice by following the temporary evolution of the separation, in
both the configuration ($\vec{r}$) and velocity ($\vec{v}$) spaces, between the
central regions of the interacting galaxies \citep[see][and references
therein]{Col14}, which in every snapshot are defined by the subsets
of the $10\%$ most bound stellar particles of each member of the
pair. We also note that by following the center of mass of a collection of particles
we avoid the relocation problems of the SMBH that arise
sometimes in simulations. All the multi-component galaxy models used in the merger experiments settle into full dynamical equilibrium in a very short time (less than one rotation). Besides we have verified that their structural and kinematic properties remain statistically unchanged for as long as a Hubble time when they are evolved in isolation.

\begin{table}
\centering
\caption{Initial orientations\tablefootmark{a} of the internal spins of progenitor galaxies.}
\label{halo_spins}

\begin{tabular}{cc}
\hline\hline

Galaxy 1 & Galaxy 2 \rule{0pt}{2ex} \\
\hline

$\odot$ & $\odot$ \\ 
$\odot$ & $\otimes$ \\
$\otimes$ & $\otimes$ \\
$\odot$ & $\downarrow$ \\
$\odot$ & $\rightarrow$ \\
$\otimes$ & $\downarrow$ \\
$\otimes$ & $\rightarrow$ \\
$\downarrow$ & $\downarrow$ \\
$\downarrow$ & $\uparrow$ \\
$\rightarrow$ & $\rightarrow$ \\
$\rightarrow$ & $\leftarrow$ \\
$\downarrow$  & $\rightarrow$ \\
\hline
\end{tabular}
\tablefoot{ 
\tablefoottext{a}{$\odot$ represents a spin oriented along the $Z+$ direction, $\otimes$ along the $Z-$ direction, $\uparrow$ along the $Y+$ direction, $\rightarrow$ along the $X+$ direction, et cetera. In all cases the orbital spin is oriented along the $Z+$ direction ($\odot$).}
}
\end{table}

The combination of the values of the parameters described above allows to build 
a total of 288 Sb+Sb
($\eta=1$) and 144 Sb+Sc ($\eta=3$) distinct merger configurations. Their evolution 
is performed using the
serial $N$-body tree-code GyrfalcON \citep{Deh00} with the adaptive
time integration scheme enabled and a longest timestep of $\sim 0.001$
simulation time units, equivalent to about $2$~Myr. This figure should
not be confused with the typical rate of the outputs of the
simulation used in the analysis, which is about one snapshot per $30$
Myr. Such rate, however, is increased to one snapshot per $\sim 10$
Myr over the $\pm 0.5$ Gyr-period around the time of coalescence of
the two nuclei to better capture the evolution of the separation of
the central regions of the galaxies during the final stages of the merger. All the
experiments begin with the galaxies separated a distance equal to 
the sum of the virial radii of their respective dark halos and are kept running 
until well after (between $\sim 1$--$2$ Gyr depending on the orbit) the formation of the merger remnant.

Regarding the merger timescale, $\tau_{\rm mer}$ -- a quantity involved in 
the calculation of some of the probabilities of detecting a DAGN outlined in
Section \ref{probabilities_def} --, it is defined as the interval
between the instant at which the center of mass of the satellite
galaxy first crosses the virial radius of the host's dark halo and the
final coalescence of the baryonic nuclei of both galaxies into a single
luminous core. Following \citet{SPV18},
the separation in phase space is quantified by the secular evolution of
the product of the (dimensionless) moduli, $\Delta r\Delta v$, of the Euclidean 
intercentric distances of the merging
galaxies in the configuration and velocity subspaces. In that same paper, especially in its Section 4, the reader can find a detailed discussion of the role played by the different parameters that configure a merger, ranging from the initial orbital circularity and energy, to the mass ratio and morphologies of the progenitor galaxies, or to the magnitude and orientation of their initial internal spins, on the length of the merger.

Since gravity is the only physics in our simulations, we could in
principle attempt to extend our predictions to different epochs
(i.e. redshifts) simply by scaling masses, times and lengths. In
practice, however, the fact that we take for both the halos'
concentration and the global properties of the stellar component values
that are characteristic of the local universe render this extrapolation 
unfeasible.

\section{Setting the scene for the triggering and detectability of DAGN}\label{DAGNmodel}

The strategy adopted for the assessment of the incidence of DAGN in the local
universe is based on the following assumptions:

\begin{itemize} 
	\item DAGN are triggered in major ($\eta\leq 3$) galaxy-galaxy mergers; 
	\item the activity of the nuclear BH is essentially encapsulated in the
      intercentric separation in phase space of the merging
      galaxies\footnote{This deliberately ignores the possible effects
        of the spin and energy of the colliding galaxies in the BH's
        accretion rate/growth.}; and
	\item the level of activity increases with decreasing intergalactic separation.
\end{itemize}

Since we are interested in comparing our predictions with a variety of
outcomes from observations and simulations, we have implemented up to
three different maxima of intergalactic separation in the phase space
when calculating the DAGN frequencies. In order to provide a sense of the
prominence of the nuclear activity, we assume that such
separations are inversely correlated in a sensible way with certain
thresholds of AGN bolometric luminosity, thus preserving the trend
that more luminous duals tend to be closer to each other
\citep{Ste16,Vol16}. The nuclear activity levels adopted are:
\begin{enumerate}[label=(\roman*)]
\itemsep0.5em
\item WEAK, usually long-term ($> 10$~Myr), DAGN activity, which we assume is related
to low bolometric luminosity thresholds of around $10^{42}$~erg~s$^{-1}$, and that we  
associate with intrinsic phase-space separations $\Delta r\lesssim
50$~kpc and $\Delta v\lesssim 200$\ \kms, in reasonable agreement with
the findings of controlled simulations \citep{VWa12,Cap17} and
observations \citep{Kos12};
\item INTERMEDIATE activity,
which is expected to occur at $\Delta r\lesssim 10$~kpc and that
likely involves bolometric luminosities $\gtrsim 10^{43}$~erg~s$^{-1}$;
and  
\item STRONG activity, triggered when $\Delta r$ becomes
smaller than about 2 kpc and where it is feasible to expect that the
typical bolometric luminosity $L_{\rm{bol}}$ of the pair can reach
values of at least $10^{44}$~erg~s$^{-1}$.
\end{enumerate}
Let us stress again that the adopted  
identifications between phase-space boundaries and nuclear activity
thresholds are only indicative and merely established to help the reader 
have a rough idea of the minimum bolometric luminosities that can be expected 
depending on the physical separation of the BH. Besides, as the luminosity of any AGN pair is expected to basically reflect the luminosity of its most massive member, we do not have accounted for the possibility that the central BH can have different masses -- something that can happen especially in unequal mass mergers -- and therefore experience different feeding rates that may lead to different luminosities.

The total lifetime of the AGN phase triggered by the
interactions (i.e. the duty-cycle of the central BH) is controlled by a parameter
$\tau_{\rm agn}$ which is allowed to range from $10$ Myr up to a
maximum of $100$ Myr \citep{Gat15,Cap17,Ble18}. The calculation of
probabilities assumes that there is no correlation between the span
and strength of AGN activity -- thus ignoring claims that AGN
lifetimes may decrease with increasing luminosity \citep{Ble18}. For
this reason, we have chosen to provide predictions for the two extreme
values of this range, which will be hereinafter referred to as the
SHORT and LONG BH duty-cycles, respectively. However, what is
explicitly taken into account in our modeling is that the nuclear
regions of the galaxies take a while to perceive the effects of the
interaction once their relative separation falls below the 
threshold adopted for the onset of a certain level of AGN activity. This happens because infalling matter must get rid of its angular momentum before it can begin to feed the central BH. This task, which is driven by kinematic viscosity and, most probably, magnetic fields, makes the accretion of matter into black holes a relatively slow process. A reasonable estimate, although admittedly crude, of the minimum time delay required for the start of any nuclear activity associated with a given intercentric distance in configuration space,
$\Delta r$, is provided by the free-fall speed in a typical disk galaxy, which has been approximated by the expression
\begin{equation}
\left[\frac{\tau_{\rm cros}}{\mbox{Gyr}}\right]\approx
0.006\,\left[\frac{\Delta r}{\mbox{kpc}}\right]\;,
\end{equation}
independently of the mass of the progenitors and the rest of
merger characteristics. Furthermore, we also assume that once the
nuclear activity of any level is triggered in a merger it will
continue unaltered as long as the conditions for the feeding of the
SMBH are met. In other words, that there is always enough material 
available to power the BH  
with an accretion rate below the Eddington limit
\citep{Kol06,She08}, so that the fuel supply does not
get substantially affected by feedback.

In an attempt to mimic the most frequent observational limitations
that are encountered when trying to determine the abundance of DAGN, we
have also implemented a procedure that aims to reproduce the incidence of dual
systems with double-peaked narrow emission lines in their optical
spectra \citep{Zho04,Ger07,Com09}. Since, as mentioned earlier, there
is a plethora of systematic surveys that apply this technique using as
parent samples different data releases from the SDSS \citep[e.g.][see
  also the references in the Introduction]{Smi10,She11,Pil12,Mul15},
we have modeled the limitations imposed by optical spectroscopic by restricting the
detections to LOS velocity differences, $\Delta v_{\rm
  1D}$, larger than $\sim 150$\ \kms\ and projected separations,
$\Delta r_{\rm 2D}$, smaller than $8$~kpc. The first constraint is set by
the resolution of SDSS spectra, while the second corresponds to the
projected distance inferred from an angle of $3$ arcsec, the diameter
of a single fiber, at a redshift of $0.15$ typical of the SDSS
Legacy Survey.

Another common approach to build up observational samples of DAGN --
which is not limited by spectral resolution or fiber size -- is to
identify them from a (ideally complete) parent dataset of bona fide
merger candidates containing individual spectral information
\citep[e.g.][]{Ell11,Kos12,Ten12,Fu15,Sat17}. It can be considered a
technique complementary of the former because it is sensitive to
galaxy pairs with nuclear separations larger than those of the
double-peak approach. To replicate what is usually done in practice, we
apply to our binary mergers up to three different filters -- in
projected intercentric distances and LOS velocities --
representative of the most typical observational constraints adopted to define
galaxy pairs in surveys at low redshift. They are:
\begin{enumerate}[label=(\roman*)]
\itemsep0.5em
\item the OPEN filter, which applies the constraints 
adopted in \citet{Liu11}, who selected pairs with $5\;\mbox{kpc}\leq
\Delta r_{\rm 2D} \leq 100$~kpc and $\Delta v_{\rm 1D} <
600$\ \kms, and where the lower limit on $\Delta r_{\rm 2D}$ is introduced to
exclude pairs in advanced mergers with nuclear separations that are
too small to be resolved by the deblending algorithm of SDSS
photometry \citep{Lup01};
\item the WIDE filter, which applies to pairs satisfying somewhat 
stricter criteria: $\Delta r_{\rm 2D}
\leq 80$~kpc and $\Delta v_{\rm 1D} < 500$\ \kms; and
\item the CLOSE filter, which is introduced to account for galaxy pairs 
selected with the conditions\footnote{$30$~kpc is also the approximate 
physical scale in projected
separation on which galaxy in pairs start to exhibit significantly
higher star formation rates than field galaxies
\citep{BGK00,Lam03,Alo04,NCA04,Per06}.} 
$\Delta r_{\rm 2D} \leq 30$~kpc and $\Delta v_{\rm 1D} <
500$\ \kms\ \citep[e.g.][]{Dar10,PA08}. 
\end{enumerate}
For the last two types of
predictions we assume that there are no special difficulties when it
comes to spatially resolve duals during the final phase of the
mergers, so we do not impose any minimum threshold in $\Delta r_{\rm
  2D}$. In any event, none of the results discussed in the following sections are  
  significantly affected by the specific values adopted for the phase-space constraints.
  
As we have just seen, the observational identification of AGN pairs
relies on projected quantities. Therefore, to derive the likelihoods
for dual-activity observability that result from any of our merger
simulations we must integrate the projections of the intrinsic
intercentric distance and velocity vectors along all possible viewing
angles. In practice, this means that we have to deal with the
cumulative distribution functions
\begin{equation}\label{F1D}
F_{\rm 1D}(\tilde{w}_{\rm 1}\leq\tilde{w}\leq \tilde{w}_{\rm
  2})=\!\int^{\tilde{w}_{\rm 2}}_{\tilde{w}_{\rm 2}}
d\tilde{w}'=\tilde{w}_{\rm 2}-\tilde{w}_{\rm 1}\;,
\end{equation}
for projections onto the 1D subspace defined by a given random LOS (it
applies to radial velocities), and
\begin{equation}\label{F2D}
F_{\rm 2D}(\tilde{w}_{\rm 1}\leq\tilde{w}\leq\tilde{w}_{\rm
  2})=\!\int^{\tilde{w}_{\rm 2}}_{\tilde{w}_{\rm 1}}
\frac{\tilde{w}'\,d\tilde{w}'}{\sqrt{1-\tilde{w}'^2}}=\sqrt{1-\tilde{w}_{\rm
    1}^2}-\sqrt{1-\tilde{w}_{\rm 2}^2}\;,
\end{equation}
for projections onto the 2D subspace defined by the plane
perpendicular to the LOS (it applies to distances in the plane of the sky), with
$\tilde{w}\equiv w/w_{\rm 3D}$ the magnitude $w$ of the projection of
an arbitrary 3D vector in units of the modulus $w_{\rm 3D}$ of the
latter, and $0\leq\tilde{w}_{\rm 1}<\tilde{w}_{\rm 2}\leq 1$. In the
present calculations no account is taken of the possibility that
attenuation by dust, and therefore viewing angle, can limit the
observable phase of AGN to a fraction of their intrinsic
lifetimes \citetext{\citealt{Hop05}; but see \citealt{Cap17}}.

Last but not least, our DAGN model includes a tunable parameter $\epsilon_{\rm agn}$ defined as
\begin{equation}
\epsilon_{\rm agn}(i) = \left\{ 
\begin{array}{ll}
0 & \mbox{if DAGN activity is not feasible,}\\
x \in (0,1] & \mbox{otherwise,}
\end{array}
\right.
\end{equation}
that measures both the effectiveness of single AGN triggering and the
simultaneity of the activity of the two central BH at each timestep $i$. It is ultimately a
measure of the detectability of correlated nuclear activity with
values ranging from 0 to 1 that encompass, respectively, the two most
extreme possibilities: i) totally ineffective triggering; and ii) fully effective triggering \emph{and} fully correlated activity. Some
hydrodynamic simulations of galaxy mergers indicate that simultaneous
BH activity requires similarly massive progenitors
\citep{BLN13,Ste16}, while others suggest that the degree of
correlation could be related to the activity strength, in the sense
that it decreases with increasing luminosity \citep[e.g.][]{VWa12}. In
any case, for the present exercise we will kept the value of this
parameter always equal to one whenever the conditions for DAGN activity are fulfilled, implying that we will be obtaining probability
estimates that operate as upper limits. Let us note that by proceeding
in this way the comparison of our outcomes with observations can be
used, for instance, to constrain the degree of correlation in the
shinning of AGN in pairs provided it is assumed that the
triggering of BH activity is highly effective.

\section{Probabilities of DAGN}\label{probabilities}

One of the most important outcomes of any investigation on the
connection between dual BH activity and mergers is the incidence of
DAGN represented by the fraction of these systems out of interacting
systems.

\subsection{Definitions}\label{probabilities_def}

Since the determination of the number of physically bound galaxy pairs
in a region or epoch of the universe is by no means simple, most
theoretical studies choose to calculate the incidence of DAGN 
by measuring instead the fraction of the total encounter time the 
pairs of active galaxies satisfy certain
conditions\footnote{This implies invoking the ergodic hypothesis and replace the ensemble average over of all the system's states in its phase space by temporal averages. In other words, assuming that by observing a binary merger for long enough time one has access to many realizations of the system. One necessary condition that must be met for this to hold is that the orbital planes and relative orientations of the galaxy pairs have to be roughly isotropically distributed \citep{KB06}.}.

This is also the view adopted in the present work. In the first place, we show 
results for two probabilities that normalize the dual activity time to
what would be the most natural measure of the length of the merger
phase, the merger timescale, $\tau_{\rm mer}$, a quantity which has a
relatively standardized definition in numerical simulations \citep[see
  e.g.][and references therein, as well as
  Sect.~\ref{experiments}]{SPV18}. They are:
\begin{enumerate} 
\itemsep0.5em
\item $\Pa$, which provides, for each and every one of our mergers, an
  estimate of the fraction of the total merger time in which dual BH
  activity, observable or not, is feasible, offering therefore a
  measure of the \emph{intrinsic} abundance of DAGN predicted by the
  major merger scenario; and
\item $\Pao$, which only considers the dual-activity time when the BH
  are observable through double-peaked narrow line features, thus providing 
  an estimate of the frequency of DAGN that could be detected 
  with optical spectroscopy in surveys free of other limitations.
\end{enumerate}
These probabilities can be expressed mathematically as:
\begin{equation}
\Pa=\frac{1}{\tau_{\rm mer}}\sum_{i=1}^{n}\epsilon_{\rm agn}(i)[t(i+1)-t(i)]\;,
\end{equation}
\begin{equation}
\Pao=\frac{1}{\tau_{\rm mer}}\sum_{i=1}^{n}p_{\rm obs}(i)\epsilon_{\rm agn}(i)[t(i+1)-t(i)]\;,
\end{equation}
where the sums go over all $n$ timesteps in which $\tau_{\rm mer}$ is divided in our simulations and where $p_{\rm obs}(i)$ measures the probability of DAGN detection calculated with the aid of Eqs.~(\ref{F1D}) and/or (\ref{F2D}) for a given set of observational constraints.

The inherent difficulties that observational studies must face when
identifying physically bound pairs of AGN, especially at large
separations, lead them to frequently constrain the characterization of
the abundance of DAGN in subsets of close pairs selected by applying
certain specific spatial and/or kinematic filters. In order to
facilitate the comparison of our predictions with this kind of
results, and with those from theoretical studies inferred along the
same lines, we have also inferred the following estimates for the fraction of DAGN 
that take into account the amount of merger time in which the intercentric
separation of the galaxies falls within the different (projected)
phase-space thresholds set out in the previous section to define galaxy pairs:
\begin{enumerate} 
\setcounter{enumi}{2}
\itemsep0.5em
\item $\Pac$, which gives the probability that a DAGN is included in a catalog of binaries; and
\item $\Paoc$, which measures the fraction of DAGN that can be
  expected to satisfy simultaneously the visibility constraints
  arising from the phase-space filtering applied for the 
  selection of galaxy pairs and the double peak-method.
\end{enumerate}
In mathematical form:
\begin{equation}
\Pac=\frac{1}{f_{\rm cat}\tau_{\rm mer}}\sum_{i=1}^{n}p_{\rm cat}(i)\epsilon_{\rm agn}(i)[t(i+1)-t(i)]\;,
\end{equation}
\begin{equation}
\Paoc=\frac{1}{\tau_{\rm mer}}\sum_{i=1}^{n}p_{\rm catobs}(i)\epsilon_{\rm agn}(i)[t(i+1)-t(i)]\;,
\end{equation}
where $f_{\rm cat}$ is the total fraction of the merger time of a bound galaxy pair during which it is expected to be included in a given catalog of binaries, while $p_{\rm cat}(i)$ and $p_{\rm catobs}(i)$ measure, respectively, the probabilities at the ith timestep that a merging system fulfills the conditions for being considered a galaxy pair and at the same time being detected as a DAGN.

\begin{figure*}
\centering
\includegraphics[width=\textwidth,height=.875\textheight,
  keepaspectratio,clip=true,angle=0]{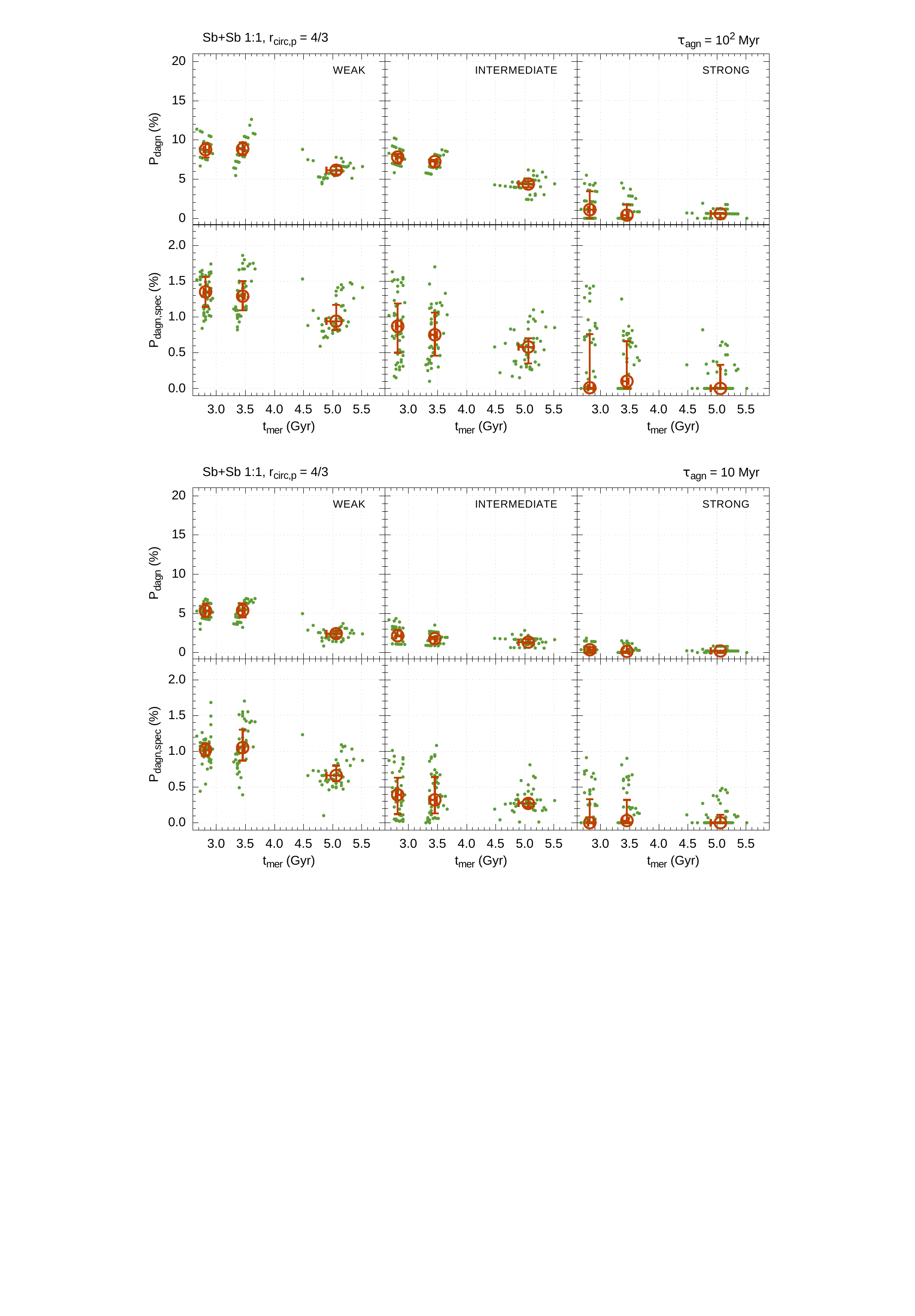}
\vspace*{-70mm}
\caption{\small Expected incidence of dual active BH
  in equal-mass mergers of spiral galaxies (of Sb type) in the nearby
  universe, as a function of the merger timescale, $\tau_{\rm
    mer}$. $\Pa$ is the intrinsic fraction of binary mergers with
  active BH pairs and $\Pao$ is the fraction of these mergers observable through
  double-peaked narrow line features in the optical window. The panels
  on each column show results for different representative thresholds
  of nuclear activity (see text). The panels in the two top rows shows results
  for a BH lifetime, $\tau_{\rm agn}$, of $10^2$ Myr, while the two bottom rows depict
  results for $\tau_{\rm agn}=10$ Myr. Individual predictions are represented by
  green dots, while large red open circles and error bars 
  show the median and interquartile range of the subsets of results
  inferred from the same initial orbital eccentricity, $\epsilon$,
  which increases from left to right in each panel. This figure is for mergers 
  starting with a reduced orbital energy $r_{\rm circ,p}=4/3$.}
\label{probsagn_spec_11_13}
\end{figure*}

The individual estimates of the above four probabilities that arise from our suite of major-merger runs are displayed graphically. The results for $\Pa$ and $\Pao$ are shown in the Figs.~\ref{probsagn_spec_11_13}--\ref{probsagn_spec_11_20} of the manuscript, while, because of their length, those for $\Pac$ and $\Paoc$ are shown in a series of plots (Figs.~\ref{probscat_spec_11_80}--\ref{probscat_spec_11_70_20}) that have been included in an Appendix of online-only material. In the panels of all these figures each single
estimate of the fractional incidence of DAGN is represented by a green dot, while the large open red symbols and associated error bars show the location (median) and scale (interquartile
range) of the subsets of results corresponding to the three different
initial orbital eccentricities considered in our merger runs. The values of these estimators are listed in Tables~\ref{probdagn} and \ref{probdagn_pair}. Note 
that the latter table only contains the probabilities for the core runs ($r_{\rm circ,p}=4/3$). The average frequencies for mergers with $r_{\rm circ,p}=2.0$ are presented exclusively in graphic form in  
Figs.~\ref{probscat_spec_11_80_20} and \ref{probscat_spec_11_70_20}.\footnote{The fact that all quoted probabilities are directly proportional to $\epsilon_{\rm agn}$ allows our results to be rescaled immediately to effectiveness of dual activity below one hundred percent or correlations less than perfect.}

\begin{table*}
\centering
\caption{Medians (M), lower (Q1) and upper (Q3) quartiles of the DAGN fraction predicted by the major merger scenario at $z\sim 0$.}
\label{probdagn}

\begin{tabular}{cccccrrr rrr}
	\hline\hline
	
	
	& & & & & \multicolumn{3}{c}{$\Pa$\tablefootmark{c}} & \multicolumn{3}{c}{$\Pao$\tablefootmark{d}} \T \\ 
	\cmidrule(lr){6-8} \cmidrule(lr){9-11} 
	$r_{\rm circ,p}$ & $\eta$ & $\tau_{\rm agn}\tablefootmark{a}$ & $L_{\rm bol}\tablefootmark{b}$ & $\epsilon$ & \multicolumn{1}{c}{M} & \multicolumn{1}{c}{Q1} & \multicolumn{1}{c}{Q3} & \multicolumn{1}{c}{M} & \multicolumn{1}{c}{Q1} & \multicolumn{1}{c}{Q3}\T \B \\\cline{1-11} 
	4/3 & 1:1 & $10^2$ & WEAK & 0.20 & 8.74 & 7.74 & 9.57 & 1.35 & 1.14 & \T  $1.56$ \\
	&  &  & & 0.45 & 8.85 & 8.04 & 9.65 & 1.29 & 1.09 & 1.50 \\
	&  &  & & 0.70 & 6.13 & 5.60 & 6.65 & 0.94 & 0.82 & 1.17 \\
	&  &  & INTERMEDIATE & 0.20 & 7.81 & 7.00 & 8.20 & 0.87 & 0.50 & 1.19  \\
	&  &  & & 0.45 & 7.22 & 6.52 & 7.48 & 0.75 & 0.46 & 1.06  \\
	&  &  & & 0.70 & 4.37 & 3.95 & 4.71 & 0.58 & 0.35 & 0.70  \\
	&  &  & STRONG & 0.20 & 1.09 & 0.00 & 3.42 & 0.01 & 0.00 & 0.76 \\
	&  &  & & 0.45 & 0.41 & 0.00 & 1.75 & 0.10 & 0.00 & 0.66 \\
	&  &  & & 0.70 & 0.58 & 0.00 & 0.67 & 0.00 & 0.00 & 0.33 \\\cline{3-11}
	&  & $10$ & WEAK & 0.20 & 5.30 & 4.52 & 6.20 & 1.02 & 0.93 & \T 1.11 \\
	&  &  & & 0.45 & 5.37 & 4.50 & 6.28 & 1.05 & 0.87 & 1.30  \\
	&  &  & & 0.70 & 2.40 & 1.89 & 2.76 & 0.66 & 0.58 & 0.80  \\
	&  &  & INTERMEDIATE & 0.20 & 2.14 & 2.07 & 2.91 & 0.39 & 0.12 & 0.63 \\
	&  &  & & 0.45 & 1.77 & 1.13 & 2.03 & 0.32 & 0.13 & 0.64 \\
	&  &  & & 0.70 & 1.32 & 1.17 & 1.72 & 0.27 & 0.22 & 0.33  \\
	&  &  & STRONG & 0.20 & 0.35 & 0.00 & 0.69 & 0.00 & 0.00 & 0.33  \\
	&  &  & & 0.45 & 0.14 & 0.00 & 0.86 & 0.03 & 0.00 & 0.32  \\
	&  &  & & 0.70 & 0.19 & 0.00 & 0.22 & 0.00 & 0.00 & 0.11  \\\cline{2-11}
	& 3:1 & $10^2$ & WEAK & 0.20 & 13.18 & 12.39 & 14.29 &  1.30 &  1.21 &  \T 1.47 \\
	&  &  & & 0.45 & 10.19 &  8.68 & 10.70 &  0.62 &  0.52 &  1.09   \\
	&  &  & & 0.70 & 8.22 &  6.85 &  9.26 &  0.86 &  0.76 &  1.02  \\
	&  &  & INTERMEDIATE & 0.20 & 8.81 &  7.75 &  9.82 &  0.97 &  0.56 &  1.39  \\
	&  &  & & 0.45 & 5.99 &  5.37 &  7.13 &  0.61 &  0.44 &  0.87  \\
	&  &  & & 0.70 & 3.59 &  3.04 &  4.21 &  0.37 &  0.31 &  0.44   \\
	&  &  & STRONG & 0.20 & 1.74 &  0.00 &  2.91 &  0.23 &  0.00 &  0.73  \\
	&  &  & & 0.45 & 1.20 &  0.00 &  2.20 &  0.00 &  0.00 &  0.50  \\
	&  &  & & 0.70 & 0.45 &  0.00 &  0.83 &  0.00 &  0.00 &  0.04  \\\cline{3-11}
	&  & $10$ & WEAK & 0.20 &  7.40 &  6.56 &  8.14 &  1.19 &  1.07 &  \T 1.38  \\
	&  &  & & 0.45 & 3.86 &  3.08 &  5.85 &  0.52 &  0.43 &  0.93  \\
	&  &  & & 0.70 & 4.23 &  3.25 &  4.59 &  0.73 &  0.65 &  0.81  \\
	&  &  & INTERMEDIATE & 0.20 & 1.24 &  1.17 &  1.98 &  0.40 &  0.19 &  0.56  \\
	&  &  & & 0.45 & 1.20 &  0.69 &  1.43 &  0.27 &  0.15 &  0.40  \\
	&  &  & & 0.70 & 1.16 &  0.85 &  1.70 &  0.22 &  0.12 &  0.30   \\
	&  &  & STRONG & 0.20 &  0.28 &  0.00 &  0.31 &  0.07 &  0.00 &  0.20  \\
	&  &  & & 0.45 & 0.20 &  0.00 &  0.23 &  0.00 &  0.00 &  0.14  \\
	&  &  & & 0.70 & 0.13 &  0.00 &  0.15 &  0.00 &  0.00 &  0.01  \\\cline{1-11}	
	2.0 & 1:1 & $10^2$ & WEAK & 0.20 & 8.33 & 7.45 & 9.18 & 1.30 & 1.18 & \T 1.43 \\
	&  &  & & 0.45 & 7.38 & 7.07 & 8.44 & 1.11 & 1.02 & 1.28   \\
	&  &  & & 0.70 & 4.25 & 3.66 & 4.78 & 0.87 & 0.76 & 0.93   \\
	&  &  & INTERMEDIATE & 0.20 & 7.30 & 6.60 & 7.65 & 0.78 & 0.52 & 1.17 \\
	&  &  & & 0.45 & 4.35 & 3.09 & 5.81 & 0.56 & 0.31 & 0.88  \\
	&  &  & & 0.70 & 2.44 & 2.06 & 2.80 & 0.41 & 0.30 & 0.52   \\
	&  &  & STRONG & 0.20 & 1.01 & 0.00 & 2.06 & 0.24 & 0.00 & 0.74  \\
	&  &  & & 0.45 & 0.73 & 0.68 & 1.49 & 0.33 & 0.00 & 0.63  \\
	&  &  & & 0.70 & 0.39 & 0.38 & 0.77 & 0.14 & 0.00 & 0.22  \\\cline{3-11}
	&  & $10$ & WEAK & 0.20 & 5.04 & 4.14 & 5.91 & 0.99 & 0.89 & \T 1.07  \\
	&  &  & & 0.45 & 4.33 & 3.38 & 5.05 & 0.89 & 0.60 & 1.04 \\
	&  &  & & 0.70 &  2.21 & 1.96 & 2.45 & 0.65 & 0.55 & 0.73 \\
	&  &  & INTERMEDIATE & 0.20 & 2.01 & 1.03 & 2.79 & 0.38 & 0.14 & 0.53   \\
	&  &  & & 0.45 &  1.25 & 0.71 & 1.48 & 0.35 & 0.11 & 0.52 \\
	&  &  & & 0.70 & 1.04 & 0.64 & 1.13 & 0.20 & 0.10 & 0.32  \\
	&  &  & STRONG & 0.20 & 0.33 & 0.00 & 0.35 & 0.08 & 0.00 & 0.25  \\
	&  &  & & 0.45 & 0.24 & 0.23 & 0.71 & 0.11 & 0.00 & 0.27  \\
	&  &  & & 0.70 & 0.13 & 0.13 & 0.20 & 0.05 & 0.00 & 0.09 \\\hline
	
\end{tabular}
\tablefoot{
\tablefoottext{a}{AGN lifetime in Myr.}
\tablefoottext{b}{Activity level/bolometric luminosity threshold (see text).} 
\tablefoottext{c}{Intrinsic.} 
\tablefoottext{d}{Observable through double-peaked narrow-line features. Probabilities are normalized to the total merger time.}
  }

\end{table*}


\subsection{Results}\label{probabilities_res}

In the next section we shall compare our data to other theoretical and
observational studies. But let us first summarize here our main
findings regarding the values and behavior of the different 
probabilities for DAGN just defined and
their dependence on the main parameters controlling the length of
mergers, namely, the initial orbital ellipticity and energy, the mass
ratio, and the spin and relative orientations of the galaxies.


\begin{figure*}
\centering
\includegraphics[width=\textwidth,height=.875\textheight,keepaspectratio,clip=true,angle=0]{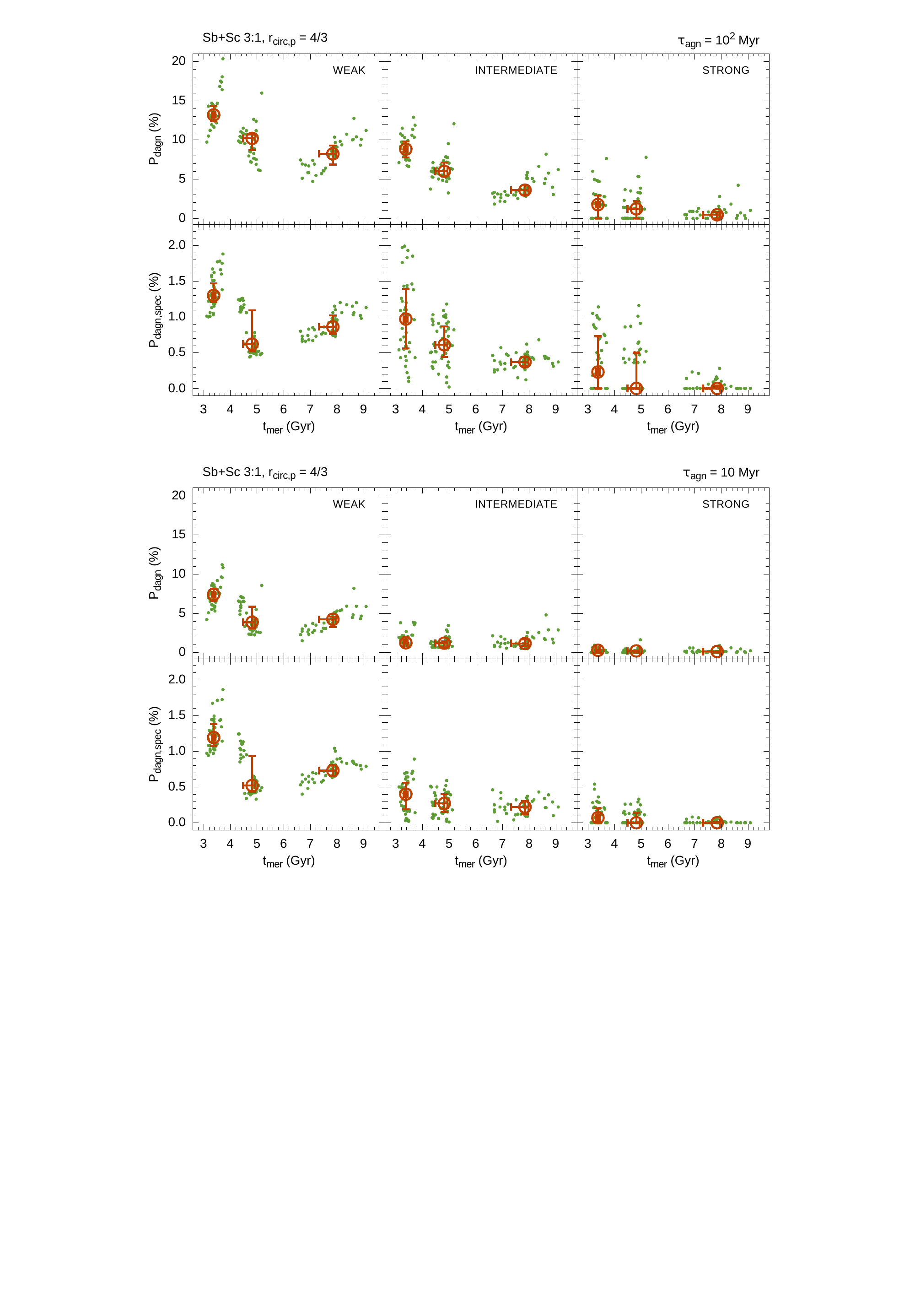}
\vspace*{-70mm}
\caption{\small Same as Fig.~\ref{probsagn_spec_11_13} but for Sb+Sc mergers
  with a mass ratio of 3:1.}
\label{probsagn_spec_31_13}
\end{figure*}

\begin{figure*}
\centering
\includegraphics[width=\textwidth,height=.85\textheight,keepaspectratio,clip=true,angle=0]{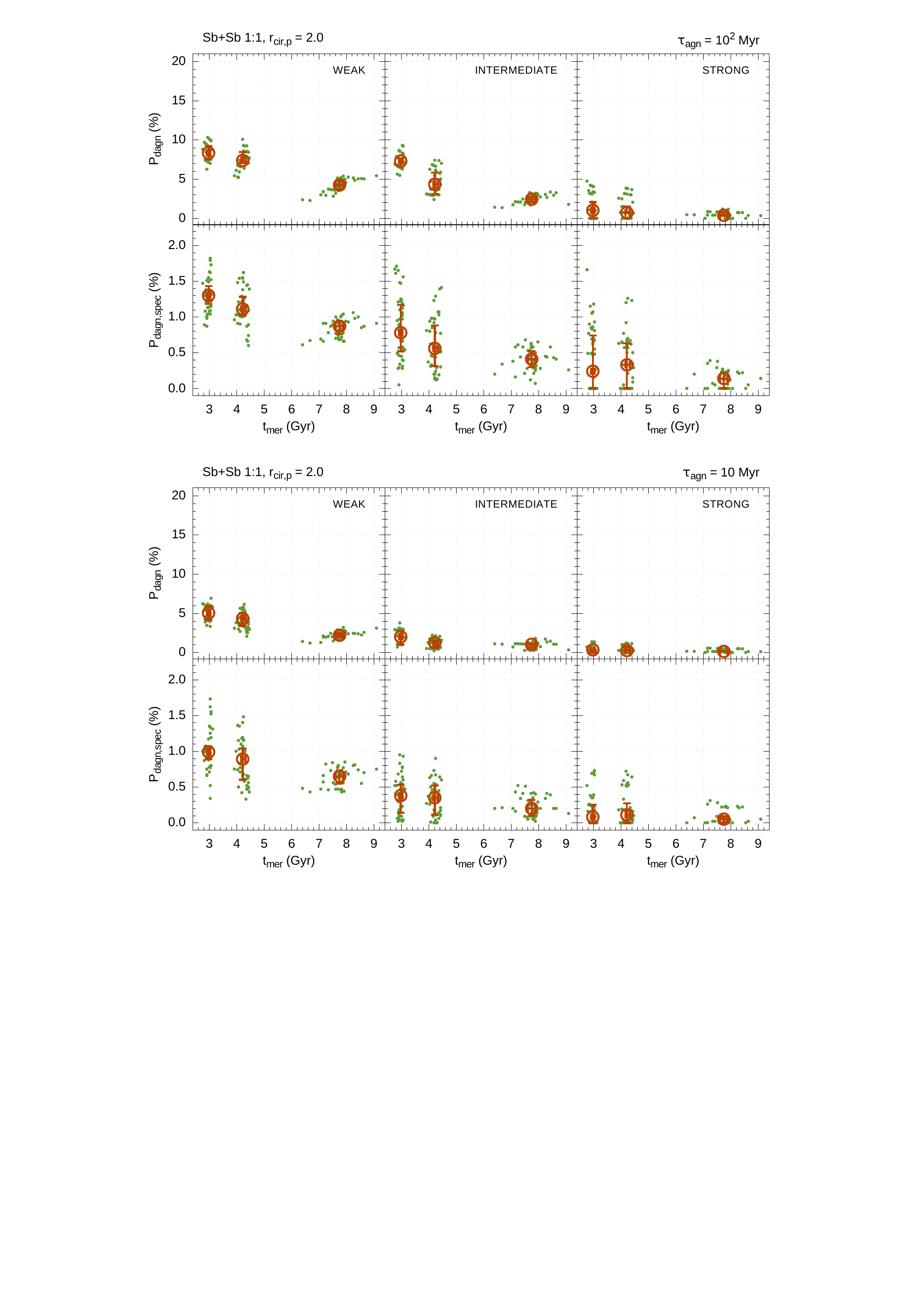}
\vspace*{-70mm}
\caption{\small Same as Fig.~\ref{probsagn_spec_11_13} but for equal-mass
  Sb+Sb mergers with an initial reduced orbital energy 
  $r_{\rm circ,p}=2.0$.}
\label{probsagn_spec_11_20}
\end{figure*}

\begin{enumerate} 
\itemsep0.5em
\item There is a mild negative correlation of $\Pa$ with the initial
  orbital circularity and the level of activity, more evident for 3:1
  mass ratios (top three panels of each six-panel group in 
  Figs.~ \ref{probsagn_spec_11_13}--\ref{probsagn_spec_11_20}). 
  For equal-mass mergers, the medians of the different
  samples of $\Pa$ inferred for the LONG BH duty-cycle
  reach up to $\sim 7$--$9\%$ for
  (moderately) radial collisions and WEAK/INTERMEDIATE luminosities,
  while for the Sb+Sc pairs this percentage raises up to $\sim 13\%$
  for the most radial and less powerful DAGN, with some individual
  predictions approaching $20\%$. As expected, the decrease of the BH duty-cycle
  reduces the values of this probability significantly, moving the
  medians towards $\sim 4$--$5\%$ for the WEAK activity regime and
  towards clearly lower frequencies for the INTERMEDIATE and STRONG
  regimes. Indeed, in this latter case, the median values of $\Pa$
  become negligible independently of the initial orbital parameters,
  the progenitors' mass ratio or the length of the BH duty-cycle.
\item Comparison of the previous results with the bottom three panels of each 
  six-panel group in Figs.~\ref{probsagn_spec_11_13}--\ref{probsagn_spec_11_20} 
  shows that $\Pao$ behaves similarly. As in the case of $\Pa$, we
  observe a weak dependency of this probability on the orbital
  characteristics, in the form of a tendency for more circular orbits
  to lead to lower typical abundances, as well as on the mass ratio,
  despite the considerable increase in the merger times that the
  latter entails. Again, the highest characteristic values of $\Pao$,
  now in the range $\sim 0.5$--$1.3\%$, are obtained for the lowest
  X-ray threshold, while not one of our hundreds of mergers leads to
  probabilities above $2\%$. 
  \vspace{2mm}
  
  In the STRONG regime (right panels of the above figures), where it is not
  surprising to find null values of $\Pao$, the spread of the
  individual predictions inferred for the LONG BH duty-cycle is
  substantial and anticorrelated with the orbital circularity: for the
  most radial orbits we record quite a number of results falling in
  the neighborhood of the interval $0.5$--$1.5\%$, while we barely
  obtain values above $0.5\%$ in the runs where $\epsilon=0.7$. Much
  like $\Pa$, we also see that the SHORT BH duty-cycle leads to a
  reduction in the location and scale of the distributions of values of $\Pao$,
  except in the case of the less powerful DAGN in 3:1 mergers, which do not
  seem particularly affected by changes in the duration of the
  periods of activity (compare the bottom left panels of the two
  six-panel groups in Fig.~\ref{probsagn_spec_31_13}). 
  This indicates that these mergers, 
  regardless of the orbital configuration, hardly lead to isolated DAGN episodes, 
  so that once the conditions for WEAK activity are satisfied they continue to hold 
  until de end of the merger process. 
\item In equal-mass mergers both $\Pa$ and $\Pao$ are independent of
  the modulus and orientation of the progenitors' halo spin. However,
  for 3:1 mergers, especially those taking place along elliptical
  orbits, these two probabilities tend to be positively correlated
  with the length of the merger phase, which in turn increases with
  the angular separation between the spin of the principal halo and
  the orbital spin \citep[see e.g.][]{SPV18} as collisions go from direct to retrograde.
\item Contrarily to $\Pa$ and $\Pao$, the probabilities normalized to
  the activity time spent within a given projected separation do not
  appear to correlate with the initial orbital circularity of the
  mergers (see Table~\ref{probdagn_pair}). As expected, 
  the more restrictive the separation criteria
  used in the normalization of $\Pac$ and $\Paoc$ the larger their
  values, while the opposite is true for both the luminosity threshold
  and the length of the activity cycle of the BH (the dependence on
  the mass ratio shows no discernible global trend). 
  \vspace{2mm}
  
  We also find that the WIDE and CLOSE filterings lead to identical results for $\Paoc$
  independently of the power emitted by the BH. This happens because
  they define pairs with no limitations in the minimum spatial
  separation, so the main restriction to the observability of DAGN
  comes from the low-velocity constraint imposed by the double-peak
  method (see the middle and right bottom panels of each six-panel group in 
  Figs.~\ref{probscat_spec_11_80}--\ref{probscat_spec_11_70_20}). 
  In contrast, both $\Pac$ and $\Paoc$
  are substantially reduced for the OPEN filter (see now the left bottom panels), 
  to the point that just in a small number
  of instances involving elongated orbital encounters and long-term
  dual activity it is feasible to first select and then observe
  double-peak narrow emission lines at peak luminosities with a non-zero
  probability. The fact that in this case we are filtering 
  out small separations makes the outcomes insensitive to AGN pairs in an
  advanced state of merger.
  \item The top six-panel groups in 
  Figs.~\ref{probscat_spec_11_80}--\ref{probscat_spec_11_70_20} 
  show that for WEAK ($L_{\rm bol}\gtrsim 10^{42}$~erg s$^{-1}$)
   DAGN with $\tau_{\rm agn}=100$ Myr the medians of $\Pac$ fall in the ranges
  $\sim 10$--$15\%$, $\sim 15$--$25\%$ and $\sim 35$--$50\%$, as
  one moves from OPEN, to WIDE and then to CLOSE separations,
  respectively (the upper limits corresponding to 3:1 mergers), while
  the change from the LONG BH duty-cycle to the SHORT one reduces them
  approximately to the half. On the other hand, $\Paoc$ shows under
  the same circumstances medians within $\sim 0.5$--$1.5\%$ for the
  WIDE and CLOSE filters and always below $0.5\%$ in the OPEN
  case. Moreover, for the most luminous DAGN we predict short
  characteristic visibility periods with all normalizations,
  particularly those corresponding to the OPEN filter, for which we
  find null medians in all cases investigated (bottom six-panel groups 
  of Figs.~\ref{probscat_spec_11_80}--\ref{probscat_spec_11_70_20}). 
  \vspace{2mm} 
  
  For its part, the dispersion of the predicted probabilities 
  tends to increase with increasing orbital
  elongation, the physical closeness of the pairs, and the progenitors mass ratio,
  being particularly sensitive to the BH lifetime (all this for the same 
  initial orbital energy; see also point $\#6$ below). We note however that, as in the case of 
  $\Pa$ and $\Pao$ for equal-mass mergers, there does not seem to be any clear 
  correlation between the individual scores of $\Pac$
  and $\Paoc$ and the internal spin of the galaxies, regarding both
  its initial magnitude and direction, and with the relative orientation of
  the latter with respect to the orbital spin \citep[see 
  also][]{Cap17}. This means that it is not easy for numerical investigations of DAGN 
  relying on idealized binary mergers to foresee a priori the minimum number 
  of orbital configurations needed to correctly represent the whole plot of possible results.
\item Changes in the initial orbital energy of mergers 
within the limits set out in Sect.~\ref{probabilities_def} essentially leave the previous conclusions unchanged (compare 
Figs.~\ref{probscat_spec_11_80} and \ref{probscat_spec_11_70} with \ref{probscat_spec_11_80_20} and \ref{probscat_spec_11_70_20}, respectively), despite corresponding to a variation of up to $50\%$.

\end{enumerate}


\section{Validation of the results}\label{comparison}

The validation of our results throughout the comparison with
observations is not at all straightforward due to the tendency of AGN
surveys to be affected by selection biases and incompleteness, the
lack of a unified estimator of the abundance of dual systems, and the
uncertainties of the measurements, which are frequently based on small
datasets. The latter problems affect simulations as well.

\begin{figure*}
\centering
\includegraphics[width=\textwidth,height=1.5\textheight,keepaspectratio,clip=true,angle=0]{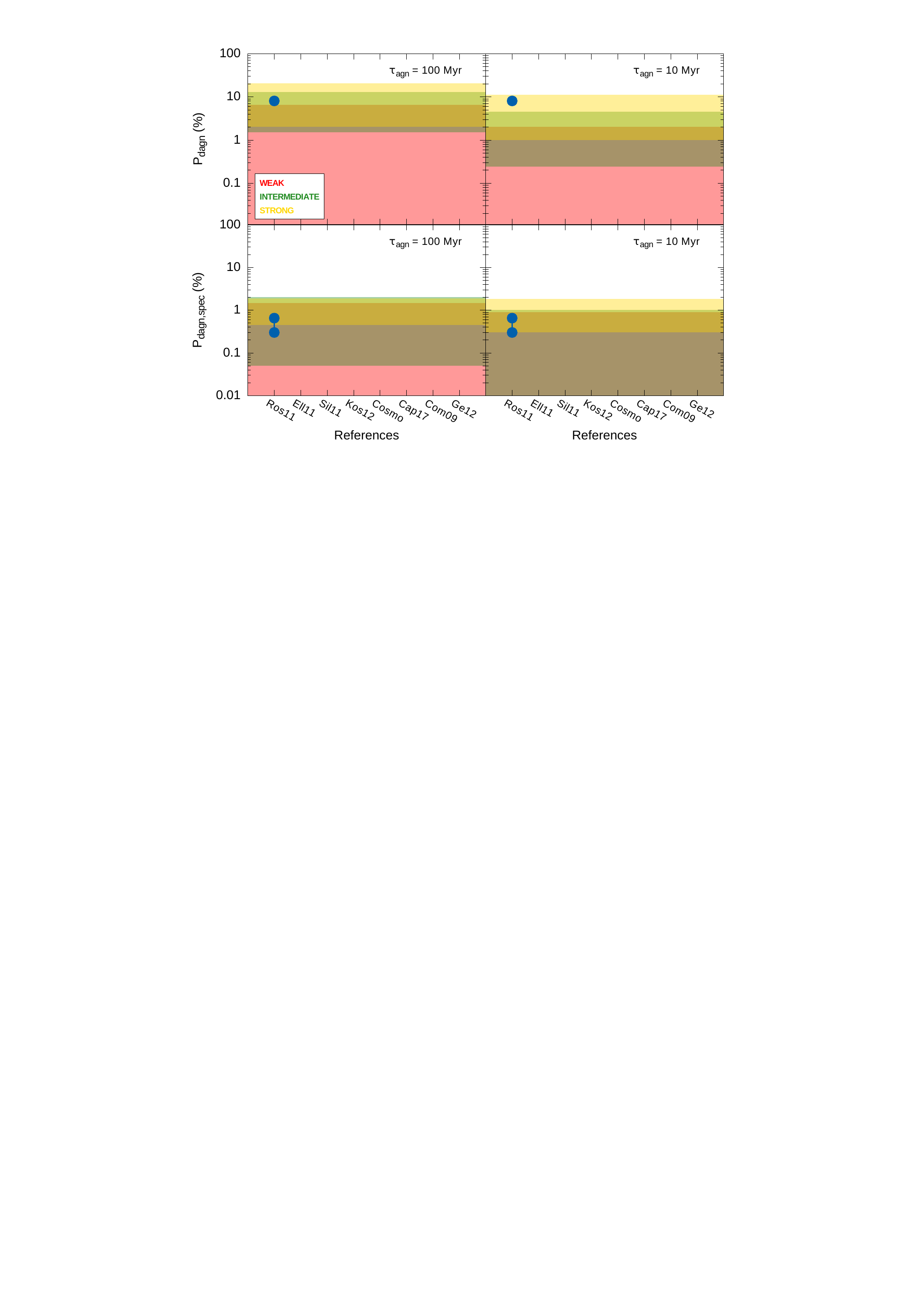}
\vspace*{-175mm} 
\caption{\small Comparison of our results for $\Pa$ (top) and $\Pao$ (bottom) with observations and other theoretical predictions. The total ranges of plausible values for these probabilities inferred from all the major mergers in our suite, i.e.\ regardless of the orbital eccentricity, orbital energy and mass ratio of the progenitor galaxies, are represented by means of horizontal bands color coded to show the different thresholds of the bolometric X-ray luminosity adopted in our numerical DAGN model (see the inset in the top-left panel). The panels on each column depict results for different durations of the AGN phase, $\tau_{\rm agn}$:  $100$ Myr (left) and $10$ Myr (right). Distributed between this figure and Fig.~\ref{rescat}, we show a set of benchmarks formed by the individual values and/or ranges of values quoted in the observational works by \citet{Ros11} (Ros11), \citet{Ell11} (Ell11), \citet{Sil11} (Sil11) and \citet{Kos12} (Kos12), which are represented by blue solid circles, those corresponding to our own estimates based on the large AGN surveys by \citet{Com09} (Com09) and \citet{Ge12} (Ge12), represented by blue solid squares, as well as those stemming from all the cosmological simulations by \citet{Ros19}, \citet{Vol16} and \citet{Ste16} (Cosmo), and from the isolated merging runs of \citet{Cap17} (Cap17), which are identified by red solid circles.} 
\label{resagn}
\end{figure*}

\begin{figure*}
\centering
\includegraphics[width=\textwidth,height=.95\textheight,keepaspectratio,clip=true,angle=0]{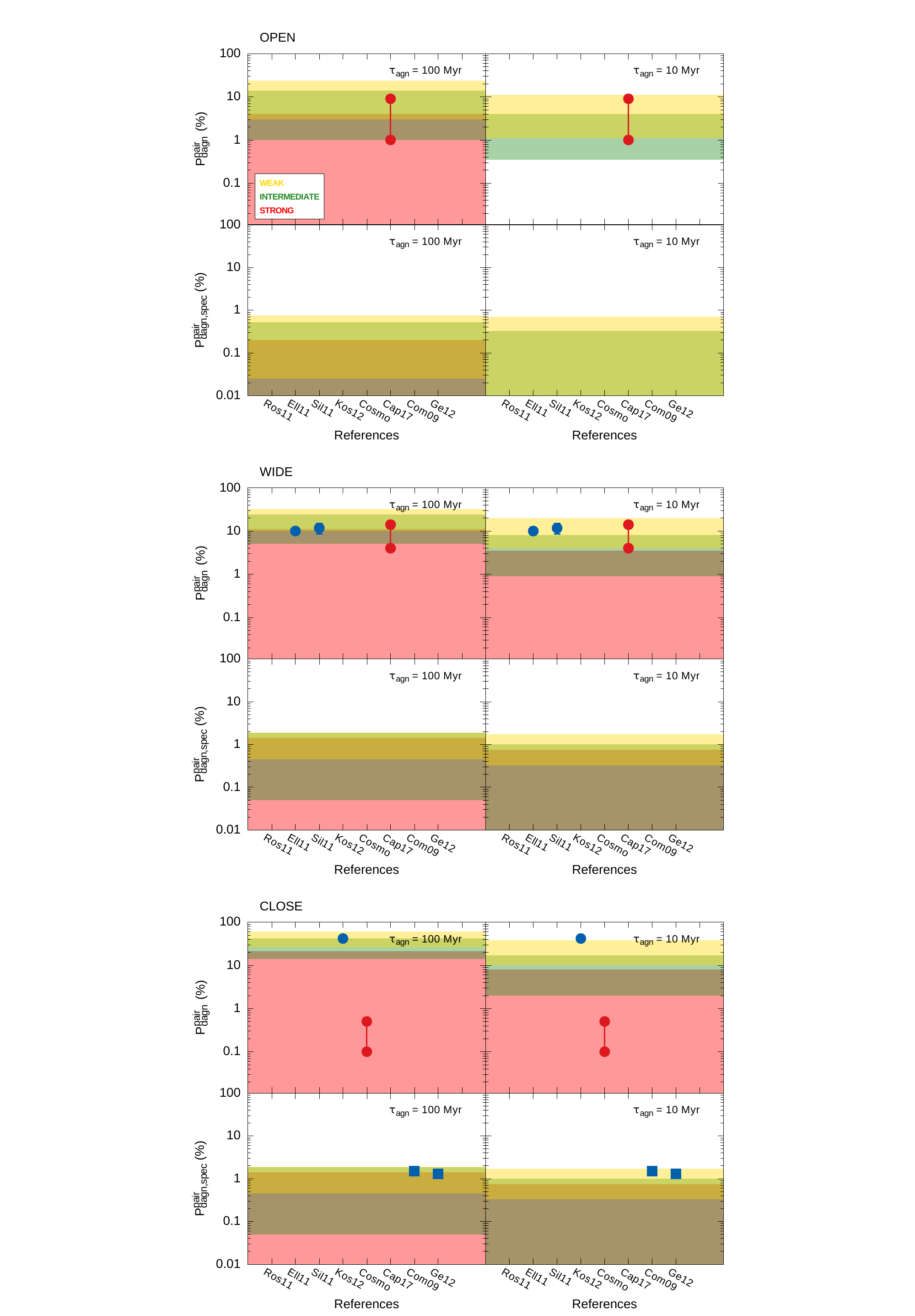}
\caption{\small Same as Fig.~\ref{resagn} but for the probabilities $\Pac$ (top) and
$\Paoc$ (bottom). The panels are shown separated into three groups, named OPEN, WIDE and CLOSE, which correspond to three different filters -- in projected intercentric distances and velocities -- representative of the most typical observational constraints adopted to define galaxy pairs in low-redshift surveys (see the text for their definitions).}
\label{rescat}
\end{figure*}

To make this section easier to read and appreciate, we have included two summary figures which provide a direct comparison of the global results of our numerical model for DAGN triggering via major mergers with the  observational results and other theoretical predictions. Fig.~\ref{resagn} shows the two probabilities that normalize the dual activity time to the total length of the merger phase, while Fig.~\ref{rescat} shows the probabilities calculated when the  normalization is the amount of merger time in which the observed intercentric separation of galaxies falls within three different phase-space thresholds.


\subsection{Comparison with observations}\label{comparison_obs}

One of the few observational studies providing measurements of the
incidence of DAGN relatively close to those defined in
Section~\ref{probabilities_def} is that by \citet{Ros11}. These
authors, starting from a small sample of (12) imaged AGN with a median
$z=0.35$, conclude that the 'global' fraction of double-peaked
emitters on kpc-scale pairs should be in the range
$0.3$--$0.65\%$. This result shows a more than fair agreement with
the total range of individual scores we obtain for $\Pao$ in our
simulations, which go from zero to $2\%$ if we assume, like they do,
that there has not been major evolution in the population of AGN
between $z\sim 0$ and $z\sim 0.4$. In addition, \citeauthor{Ros11}
apply population statistics to deliver a rough estimate of the
fraction of time a merging pair of galaxies spends in a QSO phase
under the assumption that all QSO are associated with a major merger
event. Their calculation of $8\%$ is entirely in line with the medians
of $\Pa$ we infer for the WEAK and INTERMEDIATE activity levels.

\citet{Ell11} measure directly the DAGN fraction in a large sample of
more than $11,000$ galaxy pairs extracted from the SDSS legacy
volume, where companionship is defined from the constraints $\Delta
r_{\rm 2D} < 80$~kpc and $\Delta v_{\rm LOS} < 200$\ \kms. From the
information gleaned from the \citet{Kew01} BPT classification scheme,
they find that this fraction increases steadily with decreasing
nuclear distance up to $\sim 10\%$ at the closest separations
($\Delta r_{\rm 2D} < 10$~kpc) for major pairs -- this number doubles
when they consider pairs with AGN that are either single or
double. By using our estimates of $\Pac$ for low-luminosity (WEAK)
DAGN in WIDE pairs as a proxy for this quantity (and ignoring the
differences in our respective velocity constraints), we find that the
best, and actually quite good, agreement is provided by our
predictions corresponding to the SHORT BH duty-cycle. In contrast, 
for the LONG BH duty-cycle we infer typical fractions in the range 
$\sim 15$--$25\%$ and individual values that never fall
below $12\%$. 

The results of \citet{Ell11} are corroborated by
\citet{Sil11}, who find a qualitatively similar evolution of the
fraction of moderate-luminosity ($L_{\rm bol}\sim
10^{43}$~erg~s$^{-1}$) AGN with projected physical distance using
close pairs ($\Delta r_{\rm 2D} \leq 75$~kpc and $\Delta
v_{\rm LOS} < 500$\ \kms) of massive galaxies ($M_{\rm star}> 2.5\times
10^{10}\msun$) identified in the zCOSMOS 20k catalog
\citep{Lil07}. And also similarly to the former work, they provide
measures divided according to mass ratio (though actually only for global
values because their sample is smaller). Thus, according to
\citet{Sil11}, the median fraction of galaxy pairs with a mass ratio
less than 3:1 hosting AGN is $11.7\pm 3.2$\%. Interestingly,
this value is remarkably close to the median of $\Pac$ we obtain for
WIDE pairs of INTERMEDIATE activity and, in this case, 
LONG BH duty-cycles. It
must be kept in mind, however, that this result is inconclusive, as
the observational values provided by \citet{Sil11} are likely
overestimates, because they both do not differentiate between galaxy
pairs with single or double AGN and correspond to redshifts $0.25 < z
< 1.05$ in which gas-rich mergers are expected to be considerably more
frequent than today. We also note that this survey, as the previous
one, is deficient in late-stage mergers.

\citet{Kos12} study in turn the fraction of DAGN from a sample of
167 nearby ($z<0.05$) ultra-hard X-ray-selected AGN of the all-sky
Swift Burst Alert Telescope survey. The good thing about this work is
that it not only identifies the fraction of these AGN having at least
one companion within 100 kpc, but provides detailed information on the
X-ray luminosities of the pairs, their mass ratios and their projected
separation. By examining these data (see the online-only version of
their Table 1) one can deduce, for instance, that the frequency of DAGN with
$L_{\rm bol}\sim 10^{42}$~erg~s$^{-1}$ in major pairs at projected
separations $< 30$ kpc is about $40\%$ (10/24)\footnote{It is possible
  to infer frequencies for shorter separations and/or higher
  luminosities, but this involves low number statistics and hence large
  uncertainties.}. As in the previous cases, this observational result
is also remarkably in line with our calculations, which on this
occasion are those associated with $\Pac$ for WEAK sources in CLOSE
pairs. We find the best agreement for the estimates that assume a LONG
BH duty-cycle, in which \citeauthor{Kos12}'s result fits perfectly
well, while our predictions for the SHORT BH lifetime are on average a
factor of $\sim 1.5$--$2$ smaller. Nevertheless, we caution that, as
with \citet{Sil11} work, we can only establish a comparison that is
approximated given that they consider pairs in which there is always
an active nucleus, a circumstance not included in our normalization
for $\Pac$ and that raises their estimates in an amount
difficult to gauge.

We have also attempted to extend this comparison to DAGN studies based
on samples extracted from large catalogs of individual
galaxies. This is the case, for instance, of \citet{Com09}, who
examine 1881 red galaxies from the DEEP2 Galaxy Redshift Survey, and
of \citet{Ge12}, who draw a parent AGN sample from the nearly million
objects that constitute the extragalactic spectroscopic survey of the
SDSS-DR7. In such instances, there is the possibility of using the
local value of the fraction of massive galaxies having a similarly
large companion within a projected separation of 30 kpc given in
\citet{Man12} ($0.07\pm 0.04$) to convert the fractions of
spectroscopically detected duals into frequencies that can represent a
reasonable proxy for our estimates of $\Paoc$ for CLOSE pairs and WEAK
emission (this conversion only makes sense if one assumes that major
mergers are behind all AGN fueling). The application of this simple 
re-scaling suggests that the observed fractions of spectroscopically 
detected duals in tight galaxy pairings should be around $1.5\%$ and $1.3\%$, 
respectively. These percentages, apart from being consistent with each other, 
are in very good agreement with our predictions.

\subsection{Comparison with simulations}\label{comparison_simul}

We now compare our results with estimates arising from some of the
most recent numerical studies that look into the abundance of DAGN,
either from cosmological simulations including full hydrodynamics, or
from controlled binary merger experiments like ours, but with a
explicit gaseous component.

\citet{Ros19} employ a cosmological simulation in the largest comoving
volume (100 Mpc)$^3$ from the EAGLE project \citep{Sch15} running it
up to $z=0$. They consider a visible DAGN to be an active BH pair with
a (intrinsic) separation $< 30$ kpc powering at $L_{\rm bol}\gtrsim
10^{43}$~erg~s$^{-1}$. However, instead of normalizing the
dual-activity time to the total time spent below that separation, they
choose to define the probability of detecting DAGN only in the hard X-ray bands as the average
fraction of these objects with respect to the total number of AGN and
calculate it at different cosmic epochs. At $z=0.0$--$0.5$ they find
the fraction of dual systems in which at least one of the AGN is visible in the hard X-ray band to be about $0.5\%$. As these authors point out, their
prediction is broadly consistent with similarly-defined probabilities
inferred from the outcomes of the cosmological hydrodynamic simulation
Horizon-AGN by \citet{Vol16}, and of an even larger volume simulation
included in the Magneticum Pathfinder set by \citet{Ste16}. These two
studies also find that DAGN constitute less than $0.5\%$ of all AGN, though
in the second case the estimate corresponds to $z=2$. All these
results fall considerably short of both what is inferred from
observations -- \citet{Kos12} find, for instance, that the DAGN
  fraction defined in this way, but detected using both X-ray spectroscopy and emission lines diagnostics, is $\sim 8\%$ at scales $<
  30$\ kpc -- and our estimates of $\Pac$ for INTERMEDIATE DAGN in
CLOSE pairs. If we ignore for a moment that the definition of the
fraction of visible DAGN used in \citet{Ros19} is hardly consistent
with our definition of $\Pac$, a plausible explanation for this strong
discrepancy would be the limited effectiveness of correlated nuclear
activity that characterizes AGN in cosmological simulations. According
to \citet{Ros19} there is a probability of only $3\%$ that two paired
AGN are simultaneously detected, that is to say turned on at the same
time, when they have a separation $< 30$ kpc, which they attribute to
the presence of rapid (on temporal scales of Myr) AGN variability. 
It is curious to note, however, that if we had reduced the effectiveness of
the dual nuclear activity adopted in our simulations from 1.0 to
0.03, then the bulk of our predictions for this probability would have fall between
$0.2$ and $1\%$, in much better agreement with the cosmological outcomes. 
In contrast, the comparison of our results with the 
observations carried out in Section~\ref{comparison_obs} points to effectiveness 
close to one hundred percent.

We additionally include in this appraisal the recent hydrodynamic
simulations of isolated mergers by \citet{Cap17}. These authors have
build a suite of 12 simulated mergers (6 of them major) and calculated
dual-activity observability timescales assuming different thresholds
for the bolometric luminosity, adopting different separation filters
and, no less important, translating their 3D outcomes into projected
quantities, as we have done too. In particular, in Table 2 of their paper
they list the individual frequencies for DAGN with $L_{\rm bol}\gtrsim
10^{43}$~erg~s$^{-1}$ at projected separations larger than both 1 kpc
and 10 kpc normalized to the merger time delimited by the filtering
(in an attempt to account for the constraints associated with DAGN
detection via spectroscopy they also apply a $\Delta v_{\rm LOS} \geq
150$\ \kms\ filter that, however, lacks a maximum threshold for the
projected intercentric distance, thus preventing a fair comparison with
our data). Since they are considering interacting systems with 
separations starting at $\sim 90$ kpc, it is acceptable to contrast
their figures with our estimates of $\Pac$ for, respectively, WIDE and
OPEN pairs in the INTERMEDIATE luminosity regime. At this level of activity,  
our predicted median frequencies -- taking into account the different BH duty-cycles,
mass ratios and orbital energies assumed -- range between
$\sim 4$--$16\%$ for the WIDE filter and between $\sim 0.5$--$9\%$
for the OPEN one, whereas the respective normalized times for the
simulated major encounters of \citet{Cap17} vary from $\sim 4\%$ to
$14\%$ and from $\sim 1\%$ to $9\%$. The high degree of consistency 
shown by both sets of results can be described as more than remarkable,
especially when one takes into account that our simulations do not contain an 
explicit hydrodynamic component.


\section{Summary and concluding remarks}\label{conclusions}

A physically motivated numerical model for DAGN triggering based on a large subset of the nearly six hundred collisionless simulations of major mergers presented in \citet{SPV18} has been used to predict the visibility of these systems. Our intention has been to shed light on the apparent inconsistency between what observations and theory say in this regard if, as is suspected, there is a more than probable causal connection between galaxy collisions and dual nuclear activity \citep{She10,Kos12}. The 432 bound S+S pairs selected for this task encompass a wide range of merger parameters (initial geometry and energy of the encounters, mass ratio, halos spin) covering a good number of scenarios representative of the gravitational interactions between galaxies expected to lead to DAGN activation. The ansatz at the basis of our investigation is that it is feasible to study the essential aspects of the major-merger-driven scenario for DAGN by replacing the complex gas physics involved in the fueling of nuclear activity by limits on the separation in phase space of the central regions of the colliding objects. The most outstanding feature of such treatment is that it enables the use of sets of experiments large enough to permit the statistical assessment of the effects of the parameters governing dual-activity observability, as well as the easy and intuitive inclusion of constraints in projected distance and radial velocity that mimic the most frequent limitations of AGN surveys, thus facilitating the calculation of predictions directly comparable with the existing data. The simplicity of our modeling is therefore its main strength since, at present, the realization of full hydrodynamic simulations capable of resolving in detail a similarly large number of galaxy mergers while also addressing the feeding, growing and feedback of the nuclear SMBH is still prohibitive. 

Certainly, there are also some caveats implied by our procedure, the most important being that the exclusion of the explicit treatment of the gas physics does not allow following the evolution of the AGN in a self-consistent way. This simplification hinders the applicability of our model to the study of a single encounter, but it should correctly describe, in a statistical way, the collective effects of a large number of them. Thus, it seems reasonable to expect that the results we have inferred in this work will still hold when the realization of extensive studies capable of adopting a fully realistic picture of the SMBH pairing becomes feasible. The good general agreement obtained between our predictions and the outcomes of both observations and other theoretical works can be considered as an endorsement of this expectation.

Overall, the present work shows that the inconsistency between the expected fraction of galaxy pairs undergoing synchronized nuclear activity -- inferred from arguments based on the hierarchical build-up of structure -- and the order-of-magnitude-lower abundance of spectroscopic DAGN often reported by observations \citep{Yu11} is, in a good measure, only apparent. More specifically, our calculations provide a reasonable explanation for the coexistence, in a scenario where major mergers trigger the activity of the central BH of galaxies, of theoretical predictions that place the intrinsic frequency of DAGN at levels on the order of $10\%$ \citep[e.g.][]{VHM03}, and nearby AGN surveys based on emission-line diagnostics, which systematically find fractions of double-peaked narrow-line systems at kpc-scales around $1\%$  or lower \citep[e.g.][]{Ros11}. Since the AGN phenomenon involves short-range galaxy interactions, it has also been proven that the most radical observational limitations in the detection of dual activity are those that cause a deficit in the number of very close companions (intercentric distances $\lesssim$ few kpc). 

On the other hand, our merger simulations further reveal that peak values of accretion and dual BH activity, that we tentatively associate with values of $L_{\rm bol}$ higher than $10^{44}$~erg s$^{-1}$, should be rather difficult to observe in galaxy pairs. This would not just result from the inherent difficulty that involves identifying dual systems with small spatial offsets, but also from the fact that, whatever the orbital configuration of the merger, the required physical conditions are always reached shortly before the formation of the remnant and the subsequent supermassive BH binary (a bound pair of SMBH at scales of a few pc). Our results point to intrinsic frequencies ($\Pa$) of high-luminosity DAGN that all too often fall below $1\%$, with the majority of merger configurations actually leading to null values -- the detection probabilities in close pairs, $\Pac$ and $\Paoc$, also tend to be very small, except when the most restrictive imaging filters, i.e. those with $\Delta r_{\rm 2D} \leq 30$~kpc, are applied. We also confirm, in an independent way, previous findings from both theoretical and observational studies that variations of certain factors that control the length of mergers, such as the initial orbital geometry or the mass ratio of the galaxies, can change the likelihood of DAGN detection \citep[e.g.][]{Ell11,Cap17}. Other factors, however, such as the initial energy of mergers, seem to play a secondary role. 

In addition, the fact that our experiments do not explicitly address the physics of the BH does not prevent us from drawing a few tentative conclusions in this regard through the comparison of our outcomes with those of previous works. The first has to do with the typical duration of the activity phase of the SMBH, for which we have found marginal evidence in favor of the longer-lasting periods of about 100 Myr, especially if we take into account that all our estimates are upper limits. On the other hand, we have also found indications that the activity of the central BH could be highly correlated, given that those comparisons between our estimates and other works that suggest otherwise can be attributed to significant differences in the way in which the visibility of the DAGN is defined. It is precisely this heterogeneity in establishing DAGN abundances, often conditioned by the particular characteristics of the  available data, that results in obstacles for an efficient comparison between theory and observations, or between the observations themselves, forcing us to follow a more qualitative than quantitative approach. Undoubtedly, the standardization of the measure of the DAGN fraction is something much needed in the efforts towards making a better use of the available information. However, there is little point in implementing a standard estimator of a property if it cannot be applied onto complete datasets. This is indeed the factor that most distorts the observational outcomes, as we have already mentioned in several points of this paper. Getting a complete sample of AGN is complicated since it is very difficult to quantify the biases. Even the same dataset can yield mixed results depending on the wavelength, resolution and sensitivity with which observations are conducted. In particular, there is growing evidence that black holes are likely to become heavily obscured behind merger-driven gas and dust, especially in the final merger stages when the two galactic nuclei are separated by just a few kiloparsecs \citep{Kos18}.

In summary, it has been shown that we need look no further than the most frequent photometric and spectroscopic constraints involved in the detection of DAGN to reconcile the theoretical merger rate of galaxies predicted in a hierarchical $\Lambda$CDM universe with the paucity of close AGN pairs systematically observed in the local volume. It has not been necessary to resort to the uncorrelated shining of the AGN, linked perhaps to a high variability in the accretion rates, or to a low efficiency in the triggering of the nuclear activity. In addition, no account has been taken of the many other factors that could disturb the observed frequency of AGN pairs, either by decreasing it, such as merger-driven obscuration or the tendency reported for active galaxies at small separations (i.e.\ in late-stage mergers) to be detected only in X-rays \citep{Kos12,Sat14,Ble18}, or by increasing it, such as the 'false positives' produced by double-peaked narrow-line emission associated with jets or outflows from a single AGN \citep{Sha16,Liu18}. Therefore, by reducing the tension between observations and theoretical predictions arising from the current cosmological framework, the results of the present work reinforce the support for the major merger scenario as a plausible contender among the various mechanisms that may be responsible for powering DAGN. Even so, in no way they should be taken as a confirmation that gravitational interactions, in the form of major galaxy collisions, are necessarily the single physical process capable of driving the interstellar gas to the central regions of these objects and fueling their nuclear SMBH.

\section*{Acknowledgments}

The authors acknowledge financial support from the Spanish AEI and
European FEDER funds through the coordinated research project
AYA2016-76682-C. J.M.S.\ and J.D.P., as well as I.M., extend their
gratitude to the Program for Promotion of High-Level Scientific and
Technical Research of Spain under contracts AYA2013-40609-P and
AYA2013-42227-P, respectively. C.R.A.\ acknowledges the Ram\'on y Cajal
Program of the Spanish Ministry of Economy and Competitiveness through
project RYC-2014-1577. A.d.O., I.M.\ and J.D.P.\ acknowledge financial support from the State Agency for Research of the Spanish MCIU through the "Center of Excellence Severo Ochoa" award for the Instituto de Astrof\'\i sica de Andaluc\'\i a (SEV-2017-0709)”.


\input{final_manuscript_AA_2018_33767.bbl}  

\bibliographystyle{aa}



\begin{table*}
\caption{Medians (M), lower (Q1) and upper (Q3) quartiles of the DAGN fraction in close pairs predicted by the major merger
  scenario at $z\sim 0$ for bound galaxy pairs with an initial reduced orbital
  energy $r_{\rm circ,p}$ of $4/3$.}
\label{probdagn_pair} 
\centering
\begin{tabular}{cccccrrrrrr}
\hline\hline
& & & & & \multicolumn{3}{c}{$\Pac\tablefootmark{c}$} & \multicolumn{3}{c}{$\Paoc\tablefootmark{d}$} \T \\  \cmidrule(lr){6-8}\cmidrule(lr){9-11}
$\eta$ & $\tau_{\rm agn}\tablefootmark{a}$ & $L_{\rm bol}$ & Filter\,\tablefootmark{b} & $\epsilon$ & \multicolumn{1}{c}{M} & \multicolumn{1}{c}{Q1} & \multicolumn{1}{c}{Q3} & \multicolumn{1}{c}{M} & \multicolumn{1}{c}{Q1} & \multicolumn{1}{c}{Q3}  \B \\
\hline
1:1 & $10^2$ & WEAK & OPEN & 0.20 & 9.03 & 7.63 & 10.07 & 0.18 & 0.10 & \T 0.33 \\
 &  &  & & 0.45 & 11.36 & 10.40 & 13.12 & 0.23 & 0.15 & 0.45 \\
 &  &  & & 0.70 &  8.67 & 7.65 &  9.50 & 0.26 & 0.11 & 0.36 \\
 &  &  &  WIDE & 0.20 & 16.36 &	14.70 & 18.07 & 1.31 & 1.14 & 1.51 \\
 &  & & & 0.45 & 19.26 & 17.45 & 21.02 & 1.27 & 1.02 & 1.49 \\
 &  & & & 0.70 & 15.44 & 14.50 & 16.62 & 0.94 & 0.82 & 1.17 \\
 &  &  &  CLOSE & 0.20 & 33.55 & 30.17 & 36.59 & 1.31 & 1.14 & 1.51 \\
 &  & & & 0.45 & 44.03 & 41.57 & 46.23 & 1.27 & 1.02 & 1.49 \\
 &  & & & 0.70 & 36.49 & 34.21 & 38.49 & 0.94 & 0.82 & 1.17 \\\cline{3-11}
 &  & INTERMEDIATE & OPEN & 0.20 &  8.10 &  7.52 &  8.49 & 0.12 & 0.08 & \T 0.20  \\
 &  &  & & 0.45 & 9.36 &  8.33 &  9.68 & 0.14 & 0.07 & 0.21 \\
 &  &  & & 0.70 & 6.05 &  5.17 &  6.49 & 0.10 & 0.02 & 0.22 \\
 &  &  &  WIDE & 0.20 & 14.60 &	13.19 & 14.93 & 0.83 & 0.50 & 1.14 \\
 &  & & & 0.45 & 15.46 & 14.07 & 16.17 & 0.75 & 0.46 & 1.02 \\
 &  & & & 0.70 & 11.07 & 10.17 & 12.03 & 0.58 & 0.35 & 0.70 \\
 &  &  &  CLOSE & 0.20 & 29.84 & 27.03 & 30.71 & 0.83 & 0.50 & 1.14 \\
 &  & & & 0.45 & 35.11 & 31.77 & 36.80 & 0.75 & 0.46 & 1.02 \\
 &  & & & 0.70 & 26.18 & 24.39 & 28.10 & 0.58 & 0.35 & 0.70 \\\cline{3-11} 
 &  & STRONG & OPEN & 0.20 & 0.00 & 0.00 &  1.69 & 0.00 & 0.00 & \T 0.00  \\
 &  &  & & 0.45 & 0.00 & 0.00 & 0.00 & 0.00 & 0.00 & 0.00 \\
 &  &  & & 0.70 &  0.00 & 0.00 & 0.00 & 0.00 & 0.00 & 0.00 \\
 &  &  &  WIDE & 0.20 &  2.02 &	 0.00 &  6.46 & 0.01 & 0.00 & 0.74 \\
 &  & & & 0.45 &  0.88 & 0.00 &  3.81 & 0.10 & 0.00 & 0.58 \\
 &  & & & 0.70 &  1.47 & 0.00 &  1.77 & 0.00 & 0.00 & 0.33 \\
 &  &  &  CLOSE & 0.20 & 4.09 &  0.00 & 13.51 & 0.01 & 0.00 & 0.74 \\
 &  & & & 0.45 &  2.07 & 0.00 &  9.20 & 0.10 & 0.00 & 0.58 \\
 &  & & & 0.70 &  3.39 & 0.00 &  4.12 & 0.00 & 0.00 & 0.33 \\\cline{2-11} 
 
 & $10$ & WEAK & OPEN & 0.20 & 5.45 & 4.46 & 6.12 & 0.14 & 0.09 & \T 0.28  \\
 &  &  & & 0.45 &  6.65 & 5.37 & 7.90 & 0.19 & 0.13 & 0.42 \\
 &  &  & & 0.70 &  2.61 & 1.81 & 3.45 & 0.14 & 0.05 & 0.25 \\
 &  &  &  WIDE & 0.20 & 10.07 &  8.58 & 11.76 & 0.97 & 0.88 & 1.11 \\
 &  & & & 0.45 & 11.60 & 9.84 & 13.39 & 1.03 & 0.85 & 1.30 \\
 &  & & & 0.70 &  6.18 & 4.87 &  7.06 & 0.66 & 0.58 & 0.80 \\
 &  &  &  CLOSE & 0.20 & 20.61 & 17.23 & 23.77 & 0.97 & 0.88 & 1.11 \\
 &  & & & 0.45 & 27.65 & 24.09 & 30.87 & 1.03 & 0.85 & 1.30 \\
 &  & & & 0.70 & 14.34 & 11.46 & 16.46 & 0.66 & 0.58 & 0.80 \\\cline{3-11}
 &  & INTERMEDIATE & OPEN & 0.20 &  1.85 & 1.02 & 2.21 & 0.03 & 0.02 & \T 0.08 \\
 &  &  & & 0.45 &  1.42 & 1.00 &  2.17 & 0.04 & 0.02 & 0.10 \\
 &  &  & & 0.70 &  1.31 & 0.76 &  1.76 & 0.03 & 0.00 & 0.11 \\
 &  &  &  WIDE &  0.20 & 3.98 &  3.93 &  5.54 & 0.37 & 0.12 & 0.63 \\ 
 &  & & & 0.45 &  3.82 & 2.40 &  4.33 & 0.32 & 0.13 & 0.60 \\
 &  & & & 0.70 &  3.35 & 2.95 &  4.38 & 0.27 & 0.22 & 0.33 \\
 &  &  &  CLOSE & 0.20 & 8.30 & 8.03 & 11.29 & 0.37 & 0.12 & 0.63 \\
 &  & & & 0.45 & 9.37 & 5.85 &  10.58 & 0.32 & 0.13 & 0.60 \\
 &  & & & 0.70 & 7.70 & 6.78 & 10.35 & 0.27 & 0.22 & 0.33 \\\cline{3-11} 
 &  & STRONG & OPEN & 0.20 & 0.00 & 0.00 & 0.00 & 0.00 & 0.00 & \T 0.00 \\
 &  &  & & 0.45 &  0.00 & 0.00 & 0.00 & 0.00 & 0.00 & 0.00 \\
 &  &  & & 0.70 &  0.00 & 0.00 & 0.00 & 0.00 & 0.00 & 0.00 \\
 &  &  &  WIDE & 0.20 &  0.66 & 0.00 & 1.22 & 0.00 & 0.00 & 0.29 \\
 &  & & & 0.45 &  0.25 & 0.00 & 1.81 & 0.03 & 0.00 & 0.31 \\
 &  & & & 0.70 &  0.49 & 0.00 & 0.58 & 0.00 & 0.00 & 0.11 \\
 &  &  &  CLOSE & 0.20 &  1.35 & 0.00 & 2.53 & 0.00 & 0.00 & 0.29 \\
 &  & & & 0.45 &  0.62 & 0.00 & 4.37 & 0.03 & 0.00 & 0.31 \\
 &  & & & 0.70 &  1.12 & 0.00 & 1.36 & 0.00 & 0.00 & 0.11 \\\hline
 \end{tabular}
 \end{table*}
 
\addtocounter{table}{-1}

 
\begin{table*}
\caption{continued.}
\centering
\begin{tabular}{cccccrrrrrr}
\hline\hline
& & & & & \multicolumn{3}{c}{$\Pac\tablefootmark{c}$} & \multicolumn{3}{c}{$\Paoc\tablefootmark{d}$} \T \\  \cmidrule(lr){6-8}\cmidrule(lr){9-11}
$\eta$ & $\tau_{\rm agn}\tablefootmark{a}$ & $L_{\rm bol}$ & Filter\,\tablefootmark{b} & $\epsilon$ & \multicolumn{1}{c}{M} & \multicolumn{1}{c}{Q1} & \multicolumn{1}{c}{Q3} & \multicolumn{1}{c}{M} & \multicolumn{1}{c}{Q1} & \multicolumn{1}{c}{Q3}  \B \\
\hline
 3:1 & $10^2$ & WEAK & OPEN & 0.20 & 15.62 & 14.76 & 17.17 & 0.30 & 0.25 & \T  0.44 \\
 &  &  & & 0.45 & 14.93 & 13.79 & 16.32 &  0.17 &  0.11 &  0.24 \\
 &  &  & & 0.70 & 14.21 & 10.65 & 15.42 &  0.37 &  0.29 &  0.43 \\
 &  &  &  WIDE & 0.20 & 24.08 & 22.70 & 25.27 &  1.30 &  1.21 & 1.47\\
 &  & & & 0.45 & 22.11 & 19.13 & 23.32 &  0.62 &  0.52 &  1.09\\
 &  & & & 0.70 & 21.91 & 18.23 & 24.29 &  0.86 &  0.76 &  1.02\\
 &  &  &  CLOSE & 0.20 & 43.89 & 42.15 & 45.45 &  1.30 &  1.21 & 1.47 \\
 &  & & & 0.45 & 41.02 & 32.44 & 46.88 &  0.62 &  0.52 &  1.09\\
 &  & & & 0.70 & 49.18 & 44.83 & 52.92 &  0.86 &  0.76 &  1.02\\\cline{3-11}
 &  & INTERMEDIATE & OPEN & 0.20 &  9.58 & 8.88 & 10.26 &  0.25 &  0.12 & \T 0.34 \\
 &  &  & & 0.45 &  8.56 &  7.60 &  9.25 &  0.11 &  0.06 &  0.19\\
 &  &  & & 0.70 & 4.82 &  3.33 &  5.85 &  0.11 &  0.07 &  0.17\\
 &  &  &  WIDE & 0.20 & 15.95 & 14.41 & 17.75 &  0.97 &  0.56 & 1.29\\
 &  & & & 0.45 & 13.22 & 11.99 & 15.28 &  0.61 &  0.44 &  0.86\\
 &  & & & 0.70 & 9.36 &  8.18 & 11.39 &  0.37 &  0.31 &  0.44\\
 &  &  &  CLOSE & 0.20 & 31.96 & 28.93 & 35.45 &  0.97 &  0.56 & 1.29\\
 &  & & & 0.45 & 30.48 & 27.97 & 34.83 &  0.61 &  0.44 &  0.86\\
 &  & & & 0.70 & 23.07 & 20.62 & 26.91 &  0.37 &  0.31 &  0.44\\\cline{3-11} 
 &  & STRONG & OPEN & 0.20 &  0.00 &  0.00 &  2.95 &  0.00 &  0.00 & \T  0.06 \\
 &  &  & & 0.45 &  0.00 &  0.00 &  2.33 &  0.00 &  0.00 &  0.04 \\
 &  &  & & 0.70 &  0.00 &  0.00 &  0.00 &  0.00 &  0.00 &  0.00\\
 &  &  &  WIDE & 0.20 &  3.19 &  0.00 &  5.33 &  0.23 &  0.00 & 0.70\\
 &  & & & 0.45 &  2.57 &  0.00 &  4.85 &  0.00 &  0.00 &  0.48 \\
 &  & & & 0.70 &  1.20 &  0.00 &  2.21 &  0.00 &  0.00 &  0.04\\
 &  &  &  CLOSE & 0.20 & 6.45 &  0.00 & 10.96 &  0.23 &  0.00 & 0.70\\
 &  & & & 0.45 & 6.15 &  0.00 & 12.95 &  0.00 &  0.00 &  0.48 \\
 &  & & & 0.70 & 2.96 &  0.00 &  5.53 &  0.00 &  0.00 &  0.04\\\cline{2-11} 
 
 & $10$ & WEAK & OPEN & 0.20 &  7.96 &  7.11 &  9.04 & 0.29 & 0.25 & \T 0.43 \\
 &  &  & & 0.45 &  4.70 &  3.93 &  7.62 & 0.13 & 0.07 & 0.23\\
 &  &  & & 0.70 &  6.67 &  4.94 &  7.63 & 0.34 & 0.24 & 0.39\\
 &  &  &  WIDE & 0.20 & 13.26 &  12.31 & 14.75 & 1.19 & 1.07 & 1.38\\
 &  & & & 0.45 & 8.31 &  6.69 & 12.82 & 0.52 & 0.43 & 0.93\\
 &  & & & 0.70 & 11.05 &  8.69 & 12.23 & 0.73 & 0.65 & 0.81\\
 &  &  &  CLOSE & 0.20 &  25.64 & 23.56 & 27.46 & 1.19 & 1.07 & 1.38\\
 &  & & & 0.45 & 18.38 & 14.86 & 30.26 & 0.52 & 0.43 & 0.93\\
 &  & & & 0.70 & 25.02 & 21.17 & 27.28 & 0.73 & 0.65 & 0.81\\\cline{3-11}
 &  & INTERMEDIATE & OPEN & 0.20 &  1.19 &  0.87 &  1.50 & 0.04 & 0.01 & \T  0.17\\
 &  &  & & 0.45 &  0.88 &  0.71 &  1.33 & 0.01 & 0.00 & 0.07 \\
 &  &  & & 0.70 &  0.92 &  0.71 &  1.49 & 0.04 & 0.02 & 0.11\\
 &  &  &  WIDE &  0.20 & 2.28 &  2.13 &  3.63 & 0.38 & 0.19 & 0.54 \\
 &  & & & 0.45 &  2.62 &  1.53 &  3.06 & 0.27 & 0.15 & 0.39\\
 &  & & & 0.70 &  3.11 &  2.27 &  4.58 & 0.22 & 0.12 & 0.30\\
 &  &  &  CLOSE & 0.20 & 4.65 &  4.21 &  7.35 & 0.38 & 0.19 & 0.54\\
 &  & & & 0.45 & 6.88 &  4.01 &  8.01 & 0.27 & 0.15 & 0.39\\
 &  & & & 0.70 & 7.82 &  5.71 & 10.86 & 0.22 & 0.12 & 0.30\\\cline{3-11} 
 &  & STRONG & OPEN & 0.20 &  0.00 &  0.00 &  0.00 & 0.00 & 0.00 & \T  0.00  \\
 &  &  & & 0.45 &  0.00 &  0.00 &  0.00 & 0.00 & 0.00 & 0.00 \\
 &  &  & & 0.70 &  0.00 &  0.00 &  0.00 & 0.00 & 0.00 & 0.00\\
 &  &  &  WIDE & 0.20 &  0.51 &  0.00 &  0.56 & 0.07 & 0.00 & 0.20\\
 &  & & & 0.45 &  0.42 &  0.00 &  0.51 & 0.00 & 0.00 & 0.14\\
 &  & & & 0.70 &  0.33 &  0.00 &  0.40 & 0.00 & 0.00 & 0.01\\
 &  &  &  CLOSE & 0.20 &  1.01 &  0.00 &  1.14 & 0.07 & 0.00 & 0.20 \\
 &  & & & 0.45 &  1.02 &  0.00 &  1.33 & 0.00 & 0.00 & 0.14\\
 &  & & & 0.70 &  0.77 &  0.00 &  1.00 & 0.00 & 0.00 & 0.01 \\\hline
\end{tabular}
\tablefoot{
\tablefoottext{a}{AGN lifetime in Myr.}
\tablefoottext{b}{Observational filters used to set the closeness of pairs (see text).}
\tablefoottext{c}{Measured within the phase-space filter limits.} 
\tablefoottext{d}{Simultaneously satisfying the
  constraints arising from both the filter and the double-peak method. 
  Probabilities are normalized to the merger time on filter. }
}
\end{table*}



\onecolumn

\appendix

\section{Online-only additional material}

In the online version of this article we provide additional figures
showing estimates of the incidence of DAGN in major mergers of
spiral galaxies included in local surveys of galaxy
pairs as a function of the merger timescale, $\tau_{\rm mer}$. 
In these figures, $\Pac$ is the fraction of ongoing binary mergers with active BH
pairs that can be expected in such datasets (no matter they are
observable as spectroscopic duals or not), and $\Paoc$ is the expected
fraction of these DAGN that simultaneously satisfies the condition for
detection by the double peak-method in the optical window. Each group
of six panels refers to one of the three representative levels of
nuclear activity adopted, which correspond to different thresholds of
X-ray bolometric luminosity, from top to bottom: WEAK, INTERMEDIATE and STRONG. 
Within these groups of panels, the labels OPEN,
WIDE and CLOSE refer to the criteria used in the definition of the
apparent separation of the pairs (see text). 
As in the Figs.~\ref{probsagn_spec_11_13}--\ref{probsagn_spec_11_20} of the manuscript, the green dots of the panels show the predictions derived from individual simulations, 
while the large red circular symbols and associated error bars 
illustrate the location (median) and scale (interquartile range) 
of the subset of predictions corresponding to the same initial orbital configuration, i.e.\ when only the moduli and relative orientation of the internal spins of the merging galaxies are allowed to change, whose values we collect in Table~\ref{probdagn_pair}. 


\input{figures_online_AA_2018_33767}  

\label{lastpage}
\end{document}

%% file: figures_online_AA_2018_33767.tex



\begin{figure*}
\centering
\includegraphics[width=\textwidth,height=.95\textheight,keepaspectratio,clip=true,angle=0]{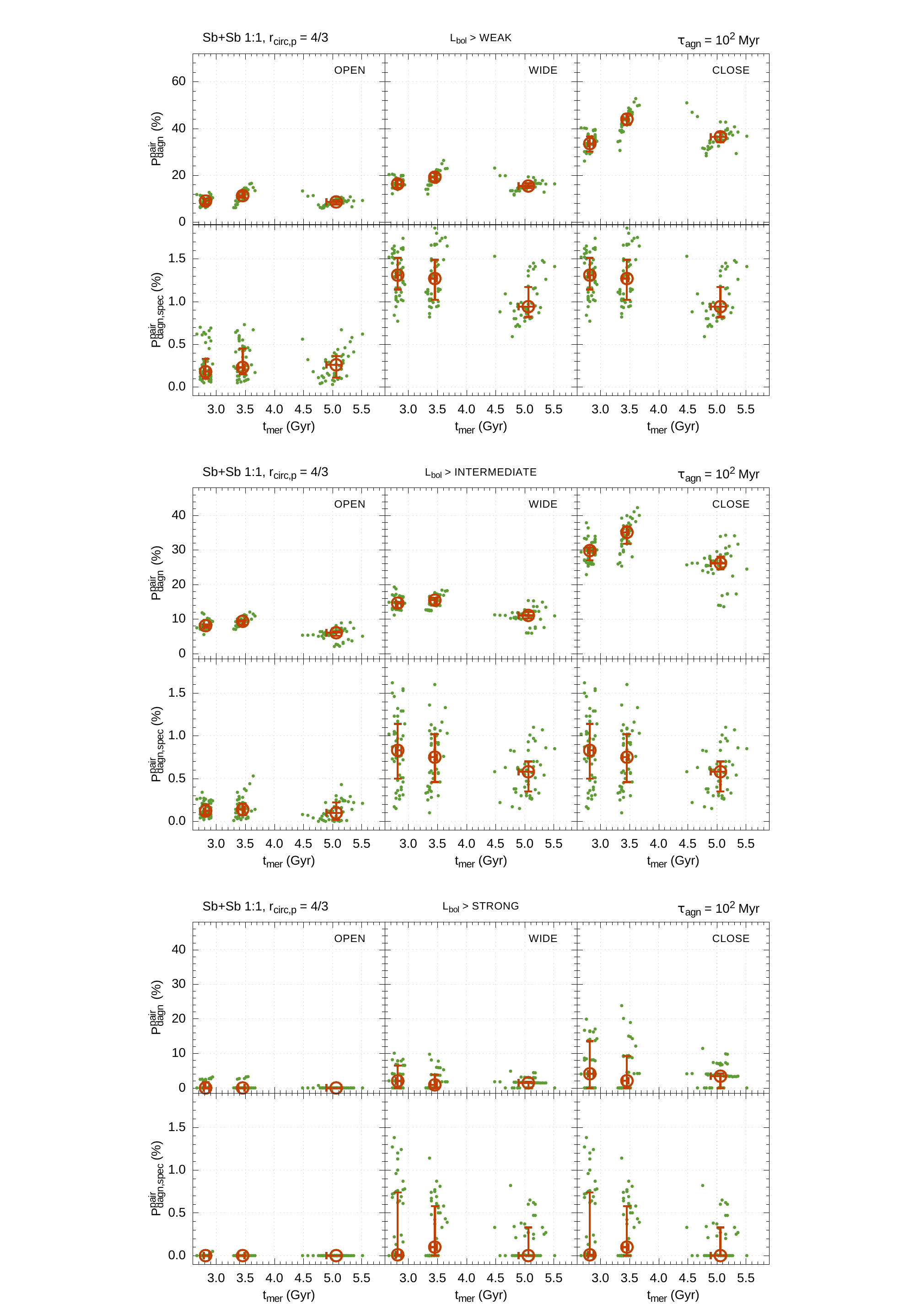}
\caption{\small Probabilities $\Pac$ and $\Paoc$ expected for equal-mass Sb+Sb mergers with an initial reduced orbital energy $r_{\rm circ,p}$ equal to $4/3$ and an AGN lifetime $\tau_{\rm AGN}$ of $10^2$ Myr.}
\label{probscat_spec_11_80}
\end{figure*}


\begin{figure*}
\centering
\includegraphics[width=\textwidth,height=.95\textheight,keepaspectratio,clip=true,angle=0]{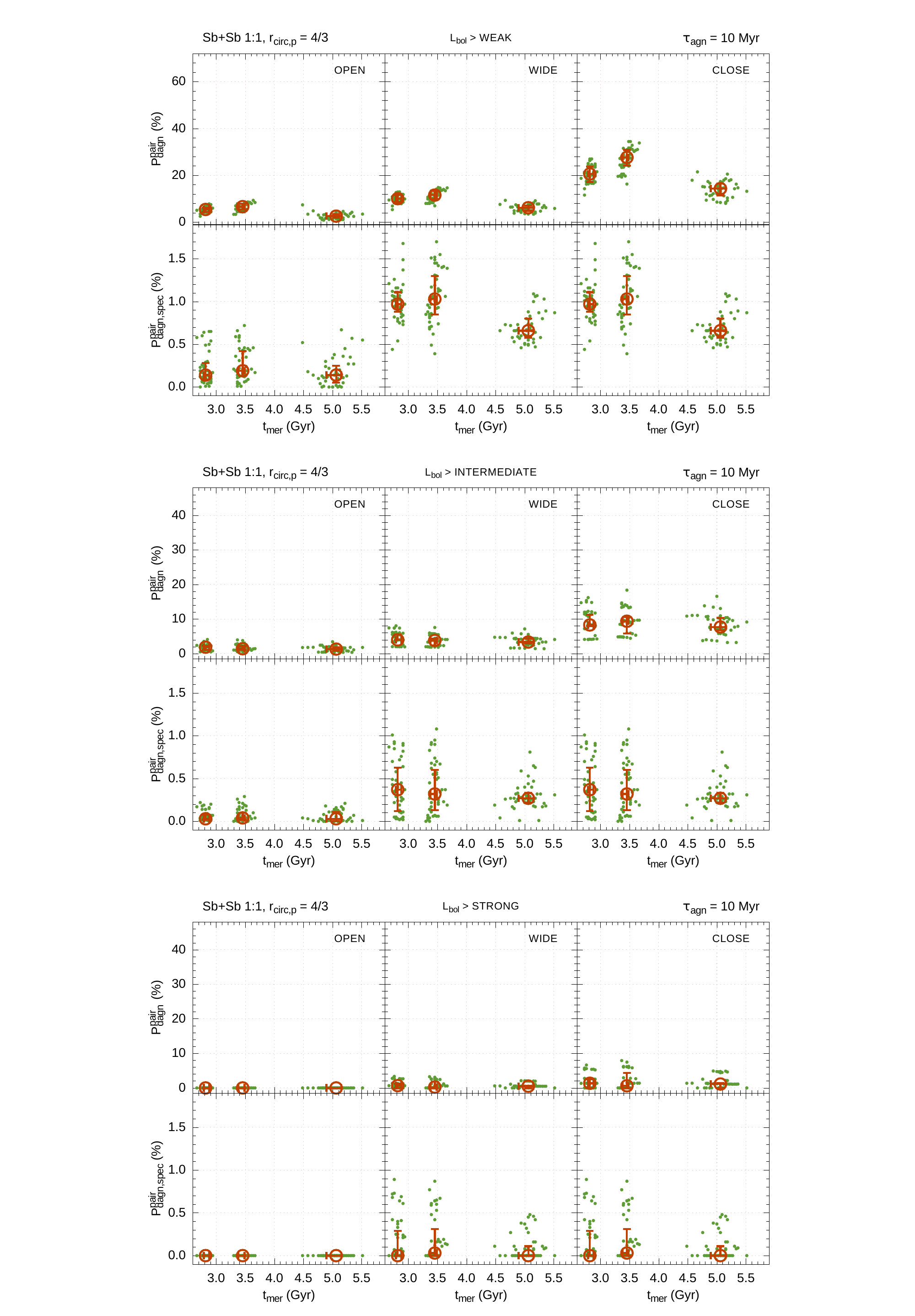}
\caption{\small Same as Fig.~\ref{probscat_spec_11_80} but for an AGN lifetime of $10$ Myr.}
\label{probscat_spec_11_70}
\end{figure*}


\begin{figure*}
\centering
\includegraphics[width=\textwidth,height=.95\textheight,keepaspectratio,clip=true,angle=0]{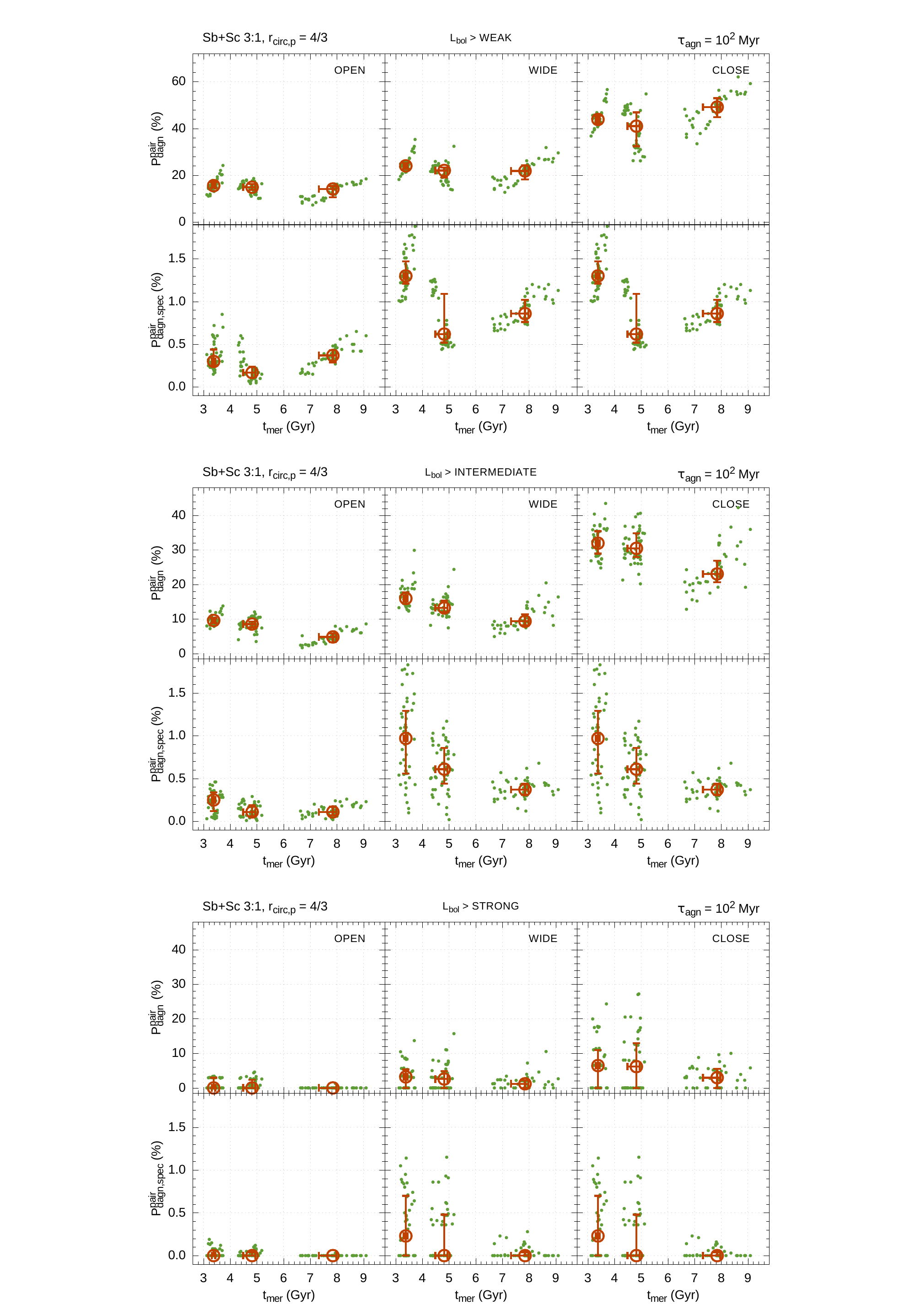}
\caption{\small Same as Fig.~\ref{probscat_spec_11_80} but for Sb+Sc mergers having a mass ratio of 3:1.}
\label{probscat_spec_31_80}
\end{figure*}


\begin{figure*}
\centering
\includegraphics[width=\textwidth,height=.95\textheight,keepaspectratio,clip=true,angle=0]{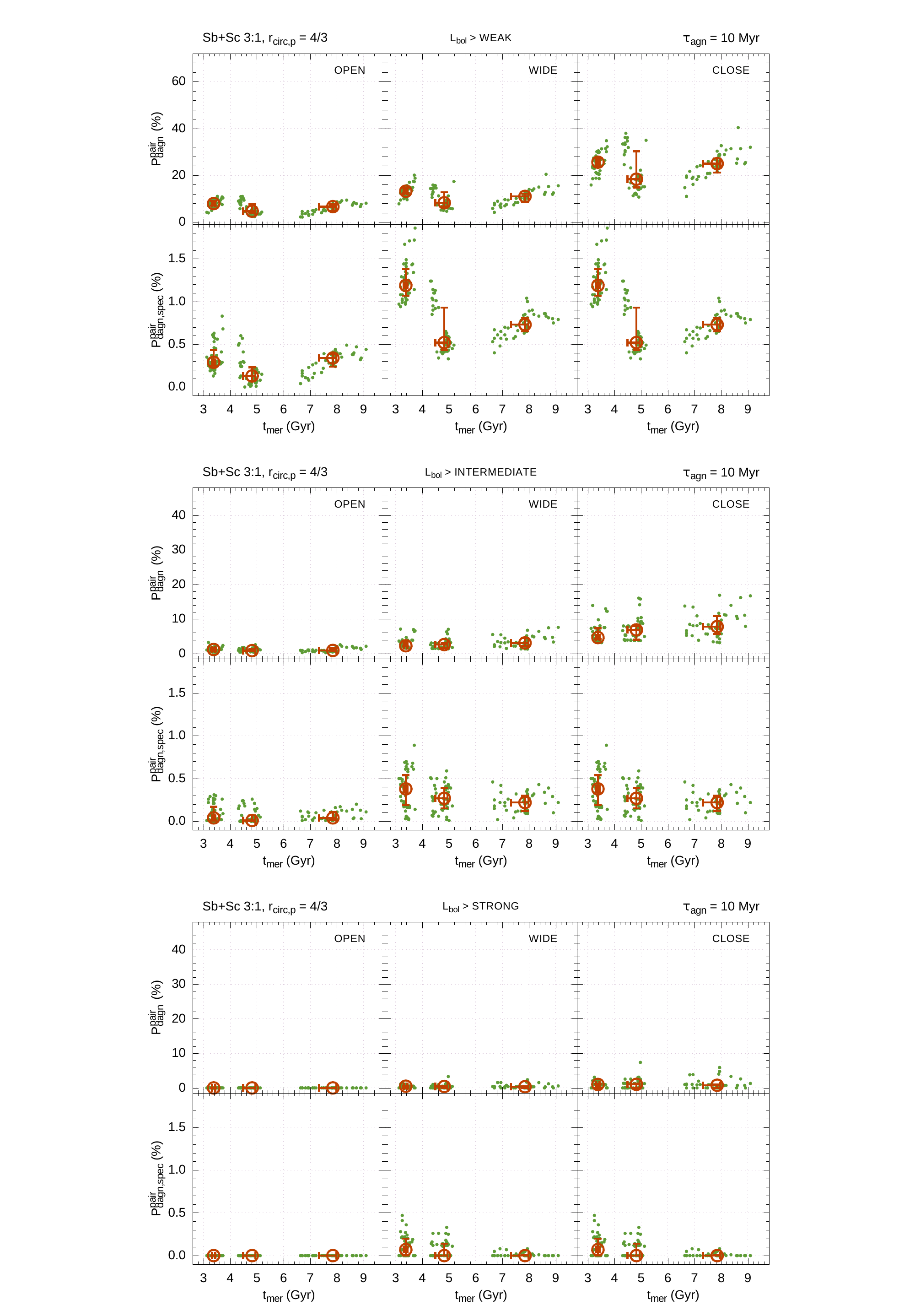}
\caption{\small Same as Fig.~\ref{probscat_spec_11_70} but for Sb+Sc mergers having a mass ratio of 3:1.}
\label{probscat_spec_31_70}
\end{figure*}


\begin{figure*}
\centering
\includegraphics[width=\textwidth,height=.95\textheight,keepaspectratio,clip=true,angle=0]{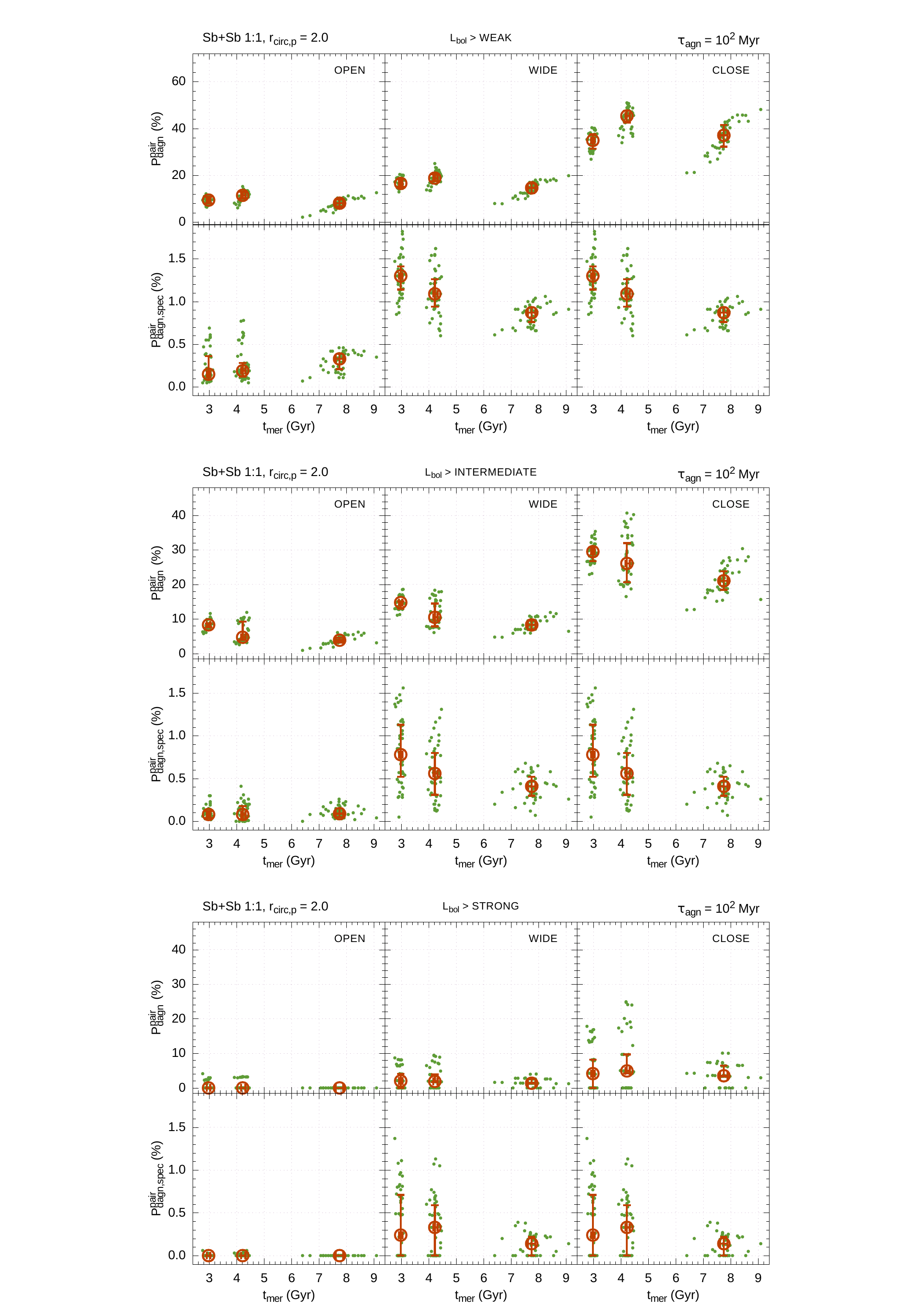}
\caption{\small Same as Fig.~\ref{probscat_spec_11_80} but for mergers with $r_{\rm circ,p}=2.0$.}
\label{probscat_spec_11_80_20}
\end{figure*}


\begin{figure*}
\centering
\includegraphics[width=\textwidth,height=.95\textheight,keepaspectratio,clip=true,angle=0]{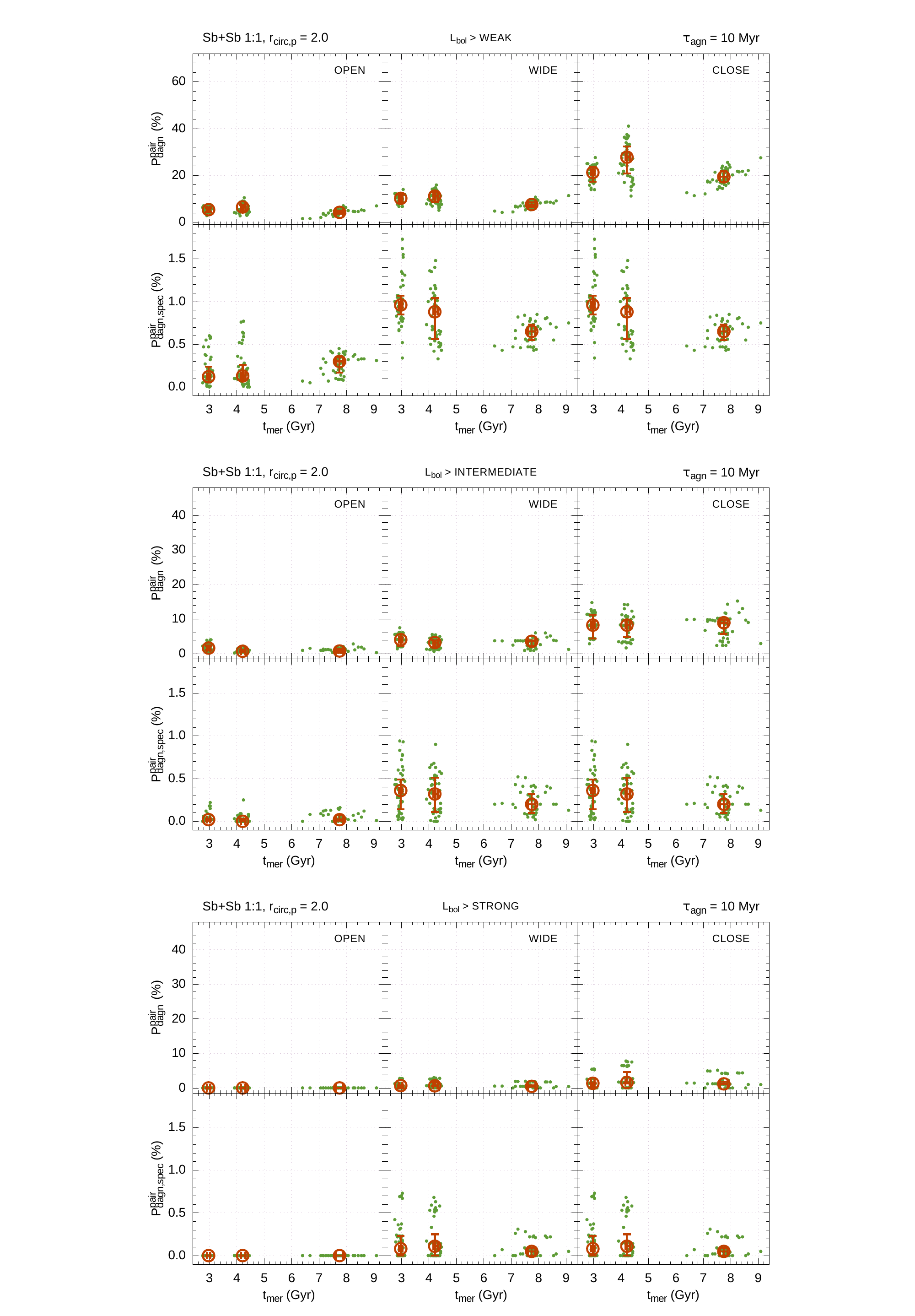}
\caption{\small Same as Fig.~\ref{probscat_spec_11_70} but for mergers with $r_{\rm circ,p}=2.0$.}
\label{probscat_spec_11_70_20}
\end{figure*}